\documentclass[11pt]{article}

\usepackage[T1]{fontenc} \usepackage[margin=1in]{geometry} \usepackage{amsmath, amssymb} \usepackage[normalem]{ulem} \usepackage{graphicx} \usepackage{booktabs} \usepackage{multirow} \usepackage{array} \usepackage{caption} \usepackage{subcaption} \usepackage{xcolor} \usepackage[round,semicolon]{natbib} \usepackage[colorlinks=true, linkcolor=blue, citecolor=blue, urlcolor=blue]{hyperref}

\newif\ifdraftmode
\draftmodefalse
\ifdraftmode
  \newcommand{\todo}[1]{\textcolor{red}{\textbf{[TODO: #1]}}}
  \newcommand{\note}[1]{\textcolor{blue}{[#1]}}
\else
  \newcommand{\todo}[1]{}
  \newcommand{\note}[1]{}
\fi
\definecolor{chengcolor}{RGB}{146,43,140}
\ifdraftmode
  \newcommand{\cheng}[1]{{\colorbox{yellow!50}{\textcolor{chengcolor}{\textbf{Comments by Dr.\ Cheng:}}}\ \textcolor{chengcolor}{#1}}}
\else
  \newcommand{\cheng}[1]{}
\fi

\graphicspath{{figures/}}

\title{Learning and Predicting Patent Technology Reuse Trajectories from Emergence-Time Signals}

\author{%
  Ayham Yousef\,$^{1}$\\ {\footnotesize ORCID 0009-0004-9698-2690}
  \and Qiang Ye\,$^{2}$\\ {\footnotesize ORCID 0000-0002-8357-221X}
  \and Qiang Cheng\,$^{1,*}$\\ {\footnotesize ORCID 0000-0002-3596-2838}
}
\date{}

\newcommand{\affiliations}{%
  \begin{center}
  \small
  $^{1}$Department of Computer Science, University of Kentucky, Lexington, KY, USA\\
  $^{2}$Department of Mathematics, University of Kentucky, Lexington, KY, USA\\[0.4em]
  $^{*}$Corresponding author. E-mail: qiangcheng8@gmail.com
  \end{center}
}

\begin{document}

\maketitle
\affiliations

\begin{abstract}
Forecasting how a newly emerged patent technology will be reused is central to technology intelligence, but reuse-pattern labels do not exist in advance: they must be constructed from the trajectories themselves, and how they are constructed determines what a forecast means. We study $201{,}710$ novel patent technologies (first-time IPC code pairings, USPTO 2002--2022). Our primary labeling applies $k$-means in the latent space of a GRU autoencoder trained on the $20$-year reuse trajectories, using no hand-crafted features; to our knowledge this is the first use of a learned sequence representation for this task. Seven emergence-time features, observable in a technology's first year, recover these labels at a one-vs-rest macro ROC-AUC of $0.914$, but the calendar year of emergence alone reaches $0.874$. A replication on technologies observed for ten full years, none of them right-censored, indicates that this calendar-year effect mainly reflects change over time in what was patented. Separately we cluster the emergence-time features themselves, after Fractal Autoencoder feature selection. That emergence-profile partition agrees with the GRU-based labels only marginally above chance (Adjusted Rand Index $\approx 0.04$), so a partition of emergence-time features is not a reuse-pattern taxonomy and should not be read as one. On a trajectory-shape task following the published construction, GBDT reaches $0.831$ with all seven features and $0.740$ with the selected subset; the published $0.728$, from a different corpus and labeling, is a reference point rather than a benchmark. Two features are additionally left-truncated for the earliest cohorts, which we quantify.

\end{abstract}

\section{Introduction}

In the rapidly evolving landscape of modern industry, tracking and predicting technological innovation trends is a critical imperative for enterprises, researchers, and policymakers alike. The United States Patent and Trademark Office (USPTO) offers an unparalleled, high-fidelity repository of historical and contemporary technological advancements. However, extracting actionable foresight from millions of patent documents presents a formidable computational challenge. Patent data is inherently large-scale, heterogeneous, and constantly expanding, making manual trend analysis impossible and traditional statistical modeling highly restrictive.

Prior research in patent analysis generally separates into two paradigms: unsupervised discovery, which clusters technologies to surface latent patterns, and supervised forecasting, which maps new technologies onto predefined categories. A key limitation in the current literature is the disconnect between them. Clustering can group technologies by behavior but does not, on its own, yield a predictor that generalizes to a newly observed technology; supervised forecasting can, but it requires labels, which are rarely available for technologies that have only just emerged. There is also little systematic evaluation of how modern machine-learning architectures, particularly deep tabular models versus traditional tree-based ensembles, perform on this kind of prediction, especially in the data-scarce regimes typical of emerging technologies.

To bridge this gap, this paper introduces a unified, end-to-end framework that constructs a reuse taxonomy without supervision and then tests how far that taxonomy is predictable from what is known at emergence. We evaluate it on a corpus of 201,710 novel patent technologies. Our primary labeling is learned from the reuse trajectories themselves: we train a GRU autoencoder to derive a latent representation of each 20-year trajectory and then cluster these representations, so a technology's label depends on how its reuse actually unfolded and on no hand-crafted quantity. We then ask whether those labels can be predicted from seven features observable in the emergence year, which is a prediction task with a genuinely external target because the labeling never saw those features. Separately, and as a second view of the same corpus, we cluster the emergence-time features directly, after applying unsupervised feature selection (FAE) to identify a compact, non-redundant subset. The two views turn out to describe different structure rather than the same structure twice, which is itself one of our findings and governs how each should be read: a partition of emergence-time features describes a technology's profile at birth, not the reuse pattern that follows.

We cast the prediction problem as a structured tabular classification task, mapping a technology's emergence-time features to its reuse-trajectory label. The forecast is made at the technology level: it characterizes the technology that a new patent introduces, which is directly useful to the firms, investors, and policymakers who must act on emerging technologies well before their long-term reuse can be observed. Given the recent resurgence of deep learning architectures tailored for tabular data, we benchmark seven distinct tabular classifiers spanning classical tree ensembles, recent deep architectures, and gradient-boosting baselines, and we evaluate them across a range of training-set sizes to probe their sensitivity to data scarcity, a frequent constraint for specialized, nascent technology sectors.

In brief, we summarize our main contributions as follows:
\begin{itemize}
    \item We develop a unified framework that couples unsupervised taxonomy construction to supervised forecasting on extensive US patent data, and we construct the taxonomy along two independent routes, so that the framework's conclusions can be checked against the choice of labeling rather than resting on it.
    \item We introduce a sequence-based labeling of reuse behavior: $k$-means in the latent space of a GRU autoencoder trained on the raw reuse trajectories, with the cluster count fixed by an elbow criterion. To our knowledge this is the first use of a learned sequence representation to cluster patent reuse trajectories. The resulting labels are recoverable from emergence-time features at a one-vs-rest macro ROC-AUC of $0.914$, and the alternative two-cluster labeling at a comparable $0.904$ binary ROC-AUC, so the finding does not rest on the cluster count. We then test how much of that predictability is temporal rather than technological, and find that emergence year alone reaches $0.874$, which bounds what the features themselves contribute.
    \item We show that the two routes agree only marginally above chance (Adjusted Rand Index $\approx 0.04$, where $0$ is the value expected of unrelated partitions), and we argue that this is a substantive result rather than a negative one: clustering emergence-time features yields a taxonomy of technologies at birth, which is not interchangeable with a taxonomy of reuse behavior, even though the two are often treated as one.
    \item We apply unsupervised feature selection (FAE) to 201,710 novel technologies and find its verdict criterion-dependent: among all 35 three-feature subsets the selected subset ranks first in Calinski--Harabasz and second in Davies--Bouldin but eleventh in the silhouette coefficient. We analyze this divergence in Section~\ref{sec:discussion} rather than report the favorable criteria alone.
    \item We establish a trend prediction benchmark by evaluating seven recent tabular classifiers (TabNet, TabM, FT-Transformer, GBDT, ExtraTrees, TabKAN, and TabMixer) on the emergence-profile labels, four of them on the independently constructed trajectory-shape task, and the two tree ensembles on the sequence-based labels, using the GBDT result reported by \citet{chen2025predicting} as a reference point rather than a like-for-like comparison.
    \item We provide an empirical evaluation of model performance under extreme data scarcity, showing that sensitivity to small training sets is architecture-specific rather than a property of model families: at the smallest training fraction ($1\%$), gradient-boosted trees degrade faster than the tree-based ExtraTrees and the best deep models (TabM, FT-Transformer), while the attention-based TabNet degrades most of all.
\end{itemize}

\section{Related Work}

\subsection{Patent Trajectory Prediction}

\paragraph{Technological trajectories.} The idea that technical change proceeds along cumulative, directed paths was introduced by \citet{nelson1977search} as natural trajectories and formalized by \citet{dosi1982technological}, who defined a technological trajectory as the pattern of problem-solving activity within a technological paradigm. \citet{sahal1985technological} described the same phenomenon as innovation avenues guided by technological guideposts. These accounts share the expectation that follow-on activity around a new technology unfolds along heterogeneous rather than uniform paths, which is the premise of sorting reuse trajectories into a small number of types. The term has since been given several empirical forms; the two most relevant here, citation-network main paths and growth curves, are reviewed below, after the literature that motivates the unit of analysis we adopt. In this paper a trajectory is the cumulative reuse curve of a single novel technology, as defined in Section~\ref{sec:setup}.

\paragraph{Invention as recombination.} Our definition of a novel technology, the first co-occurrence of a pair of IPC codes, rests on the view that invention is the recombination of existing components \citep{weitzman1998recombinant, arthur2007structure}. \citet{fleming2001recombinant} develops that view for patents, treating inventive activity as search over combinations of technology classes whose payoff is uncertain until a pairing is actually tried, and \citet{uzzi2013atypical} show that in science the highest-impact papers pair conventional combinations with a tail of atypical ones. This literature accounts for why new combinations arise and judges their value once their consequences are visible, whereas our task is prospective: predicting the shape of a pairing's reuse over the following two decades from what is observable in its first year. We use the word reuse throughout in the narrow sense of a later invention applying an existing code pairing.

\paragraph{Measuring novelty from patent codes.} Empirically, technological novelty is most often measured through patent classification codes. \citet{strumsky2012patent} use technology codes and their new combinations to trace technological change, and \citet{strumsky2015identifying} separate novelty that arises from entirely new codes from novelty that arises from first-time combinations of existing codes. \citet{verhoeven2016measuring} extend this line by combining novelty in recombination with novelty in knowledge origins into a set of patent-based indicators. A natural follow-up question is whether such measures anticipate what a patent later becomes, and \citet{kim2016technological} report that a patent's novelty profile predicts its future citation impact. These indicators are computed at the moment a patent appears, and they compress its later significance into a single quantity, typically a citation count. We retain the emergence-time perspective but change the target of prediction: we use emergence-time descriptors of a newly appeared technology to predict the shape of its reuse over the following twenty years, so that the distinction between the emergence of a technology and the trajectory along which it is subsequently reused becomes the object of prediction rather than a scalar score.

\paragraph{Follow-on use as an indicator of importance.} Counts of follow-on use are the standard indicator of an invention's technological importance. Forward citation rates were validated early as a marker of technologically important patents \citep{carpenter1981citation}, and citation-based measures of invention value date back at least to \citet{trajtenberg1990penny}. A large indicator literature has since been built on that foundation, and \citet{arts2013inventions} test whether those indicators actually identify the inventions that shaped a field's trajectory, finding that ex post impact indicators outperform ex ante novelty indicators and that combining several indicators gives the most comprehensive picture. That contrast is what motivates prediction rather than measurement: impact indicators are only available once the impact has accumulated, so an indicator that is useful at the moment a technology emerges has to be built from signals observable in its first year. Our task belongs to this family, with one difference: the predicted outcome is the shape of the whole reuse trajectory rather than a count at a fixed horizon.

\paragraph{Trajectory shape and delayed recognition.} The shape of a trajectory carries information that its endpoint count does not. Highly novel work often experiences delayed recognition, with impact accruing over long windows rather than in the years immediately following its appearance \citep{wang2017bias}. The same delay is visible in patent reuse, where some patents remain dormant for years before their citations rise sharply \citep{hou2019patent}. Under such patterns two technologies can reach similar twenty-year totals by very different paths. This motivates clustering entire trajectories; a threshold on a cumulative count would discard the shape. It also bears on the interpretation of the right-censored cluster we report in Section~\ref{sec:results-clustering}.

\paragraph{Mapping trajectories from citation networks.} The dominant scientometric operationalization of a technological trajectory is the main path through a citation network, introduced by \citet{hummon1989connectivity} and applied to patents by \citet{verspagen2007mapping}. Later work has refined how these paths are constructed, including corrections for the citation time lags that would otherwise exclude recently granted patents from the network \citep{hwang2019extending}. What such methods recover is nonetheless a retrospective account of how a field developed, traced through citations that accumulate for years after the inventions in question. The reuse trajectories we study are a different object, the cumulative follow-on inventions of one novel technology, and our task is to predict their class at emergence, from information available in the technology's first year.

\paragraph{Growth curves and technology life cycles.} A second operationalization fits parametric S-shaped curves to cumulative adoption or patent counts. \citet{bass1969newproduct} formalized the diffusion of a new product as a growth curve whose parameters separate innovation from imitation, and that functional form was subsequently carried onto patent data over long historical spans: \citet{andersen1998evolution} traced the evolution of technological trajectories across US patent classes from 1890 to 1990, and \citet{andersen1999hunt} fitted logistic growth functions to cumulative patent stocks by technology group in an explicit search for S-shaped growth paths. A related line replaces the continuous curve with a small number of discrete life-cycle stages, from emergence to decline, and infers the current stage from the systematic variation of patent indicators across those stages \citep{haupt2007patent}. Both variants fix their object in advance, either as a common functional form or as an ordered sequence of stages. The reuse trajectories we study are a closely related object, the cumulative reuse of one code pairing rather than of a field, but we treat them without a parametric form: cluster prototypes take the place of fitted curve families, and the classes are learned from the data.

\paragraph{Data-driven forecasting from patent data.} Forecasting technology emergence and success from patent and bibliometric data has a long history. Early work mined the text of scientific and patent records to predict which technologies were about to emerge \citep{smalheiser2001predicting}. A later line of research shifted from textual signals to quantitative patent indicators, forecasting the eventual success of a technology from indicators computed on patent records \citep{altuntas2015forecasting}. More recent work brings deep and language-model representations to the same problem, identifying emerging technology topics from the multi-field characteristics of patented inventions \citep{song2023identification}. Across these successive generations of methods the prediction target has changed little: whether a field or topic will emerge, or whether a technology will reach some scalar measure of success. The multi-year reuse trajectory of an individual novel technology, and in particular the shape that trajectory takes over its full lifetime, is a less common target.

\paragraph{Cluster-then-classify designs and the present work.} Deriving labels by unsupervised clustering and then training a supervised model to predict them is an established design in patent analytics. \citet{kyebambe2017forecasting} cluster patents and train supervised classifiers on the automatically labeled clusters, and \citet{lee2025dynamic} group patent-indicator time series with dynamic time warping to identify diffusion patterns. We build most directly on \citet{chen2025predicting}, who define a set of emergence-time features for novel patent technologies, cluster their reuse trajectories by shape using dynamic time warping \citep{berndt1994dtw}, and then train a gradient-boosted classifier to predict cluster membership, using a SHAP analysis to identify \texttt{INVENT\_APPL} as the most important predictor. The same trajectory construction underlies \citet{pezzoni2022fast}, who define a novel technology as the first-ever co-occurrence of two IPC codes and study the stream of follow-on inventions that reuse it, modeling antecedents of follow-on counts and not grouping trajectories by shape. We adopt the feature definitions of \citet{chen2025predicting}, and their reported ROC-AUC serves as a reference point. Where they derive reuse-pattern labels by DTW-based $k$-means on the raw trajectories, we derive labels by two further unsupervised routes, one over a learned sequence representation and one over selected emergence-time features, and we measure directly how far the resulting labelings agree (Sections~\ref{sec:results-transfer} and~\ref{sec:results-gru}).

\subsection{Learned Representations for Sequence Clustering}
\label{sec:related-seqclust}
Clustering of time series is conventionally divided into raw-data-based, feature-based, and model-based approaches \citep{liao2005clustering, aghabozorgi2015time}. The emergence-profile route of this paper is not a time-series clustering at all: it partitions technologies on covariates observed at emergence that are never computed from the trajectory. The sequence-based route is, and within that literature it sits on the learned-representation side of a specific divide. Raw-data (shape-based) methods compare sequences under a distance fixed in advance, usually an elastic one such as dynamic time warping \citep{berndt1994dtw}; barycenter averaging made DTW $k$-means practical \citep{petitjean2011global}, and fuzzy \citep{izakian2015fuzzy} and other shape-based variants \citep{li2022new} followed. Fixing the distance fixes what counts as similar before any data is seen, which for reuse trajectories means committing in advance to whether two curves that rise at different rates are the same shape.

The alternative is to learn the representation and cluster in it, the approach taken by deep clustering \citep{min2018survey}, of which the best known instance alternates between a learned embedding and a cluster assignment that sharpens it \citep{xie2016unsupervised}. We use the simpler and more conservative form of this idea: train a sequence autoencoder to reconstruct the trajectories, then cluster the resulting codes once, without any clustering term in the training objective. This keeps the representation answerable only to reconstruction, so the partition cannot be an artifact of a clustering loss that was itself optimized. Sequence autoencoders of this kind have proved effective for molecular representation learning \citep{winter2019learning, mucllari2023novel}. The components are therefore standard; what is new here is the object they are applied to and the comparison they make possible. To our knowledge no prior work clusters patent reuse trajectories in a learned sequence representation, and none measures how far such a labeling agrees with a labeling built from emergence-time features on the same corpus.

\subsection{Unsupervised Feature Selection}
Unsupervised feature selection methods generally fall into three families. Filter methods score each feature on its own, independent of any downstream model, for example by how well it preserves local structure in the Laplacian Score \citep{he2005laplacian}. Spectral and embedding methods such as MCFS \citep{cai2010mcfs} instead select features that preserve cluster structure in a learned low-dimensional space. Embedded autoencoder methods learn the selection jointly with reconstruction, as in the Concrete Autoencoder \citep{abid2019concrete}, which makes a discrete feature mask differentiable. The Fractal Autoencoder (FAE) of \citet{wu2021fractal}, which we adopt, belongs to this third family: it pairs a global reconstruction objective with a sub-network that must reconstruct all features from only the top-$K$ selected ones, regularized by an $\ell_1$ penalty on the learned importance weights (detailed in Section~\ref{sec:method-fae}). We prefer FAE because the sub-network constraint rewards a subset that is representative as a whole, which suits a small and correlated feature set of the kind we use. What all three families share is that they return a subset of the original variables, and that property is what our setting demands: the retained emergence-time features must remain interpretable as named quantities, so non-linear dimension reduction, which yields new coordinates rather than a subset, is not an appropriate substitute.

\subsection{Tabular Deep Learning}
A good deal of recent work asks whether neural architectures can match the accuracy that tree ensembles reach on tabular data. TabNet \citep{arik2021tabnet} applies sequential attention over features to select which inputs each decision step uses; FT-Transformer \citep{gorishniy2021revisiting} adapts the transformer encoder to mixed numerical and categorical inputs; and TabM \citep{gorishniy2025tabm} uses parameter-efficient model ensembling to improve robustness. More recent architectures include TabKAN \citep{eslamian2025tabkan}, which replaces conventional layers with Kolmogorov--Arnold networks whose learned univariate functions afford a degree of interpretability, and TabMixer \citep{eslamian2025tabmixer}, an MLP-mixer variant that forgoes attention and convolution for efficiency. Extremely Randomized Trees \citep{geurts2006extratrees}, which add randomization to the split selection of standard ensembles, supply a tree-based point of comparison. This literature evaluates such models on established benchmark tasks, where the labels are given and the feature set is fixed. We instead apply them to labels obtained by unsupervised clustering, using only the features observable in a technology's first year, and evaluate them under two independently constructed labelings so that no conclusion rests on a single labeling procedure.

\section{Methodology}
\label{sec:method}

Our framework predicts the reuse pattern of a novel technology from features available at its emergence. Because no reuse labels exist in advance, the labels must themselves be constructed, and we construct them along two separately built routes. The primary, sequence-based route clusters the reuse trajectories in a learned representation and uses no hand-crafted features (Section~\ref{sec:method-gru}). The second route selects among the emergence-time features and clusters those (Section~\ref{sec:method-fae}). A supervised stage then predicts each labeling from the emergence-time features. We then describe validation procedures that test whether the learned classifiers reflect genuine structure rather than an artifact of either labeling.

\subsection{Problem Formulation}
\label{sec:method-problem}
Let $\mathbf{X} \in \mathbb{R}^{n \times m}$ denote the feature matrix for $n$ novel technologies described by $m=7$ features (Section~\ref{sec:setup}). Each technology $i$ also has a reuse trajectory $\mathbf{t}_i \in \mathbb{R}^{W}$, the cumulative count of subsequent inventions over a $W=20$ year window. The trajectories encode the eventual reuse behavior we wish to characterize, but they are observable only in hindsight; the features are available at emergence. Our goal is to learn a mapping $\hat{y} : \mathbb{R}^{m} \to \{1,\dots,k\}$ from emergence-time features to a reuse-pattern label, so that the pattern of an unseen technology can be predicted early.

Because the data carries no ground-truth pattern labels, the labels must themselves be induced, and the choice of how to induce them is a modeling decision rather than a preliminary. We therefore construct labels along two routes and keep them distinct throughout. The primary route (Section~\ref{sec:method-gru}) clusters the trajectories in a learned sequence representation, so that a label reports how reuse actually unfolded; predicting such a label from $\mathbf{X}$ is a genuine prediction problem, because the labeling never observes $\mathbf{X}$. The second route (Section~\ref{sec:method-fae}) clusters the emergence-time features themselves and so describes a technology's profile at birth; recovering such a label from $\mathbf{X}$ is a consistency check, not a forecast, since target and input share the same information. We use the term \emph{reuse-trajectory label} for the first and \emph{emergence-profile label} for the second; where the contrast between the two routes is what matters we also call them the sequence-based and the feature-based route, which name the same two objects. Section~\ref{sec:method-validation} sets out how each is validated and how far the two agree.

\subsection{Sequence-Based Labeling via a Recurrent Autoencoder}
\label{sec:method-gru}
\begin{figure}[htbp]
  \centering
  \includegraphics[width=\linewidth]{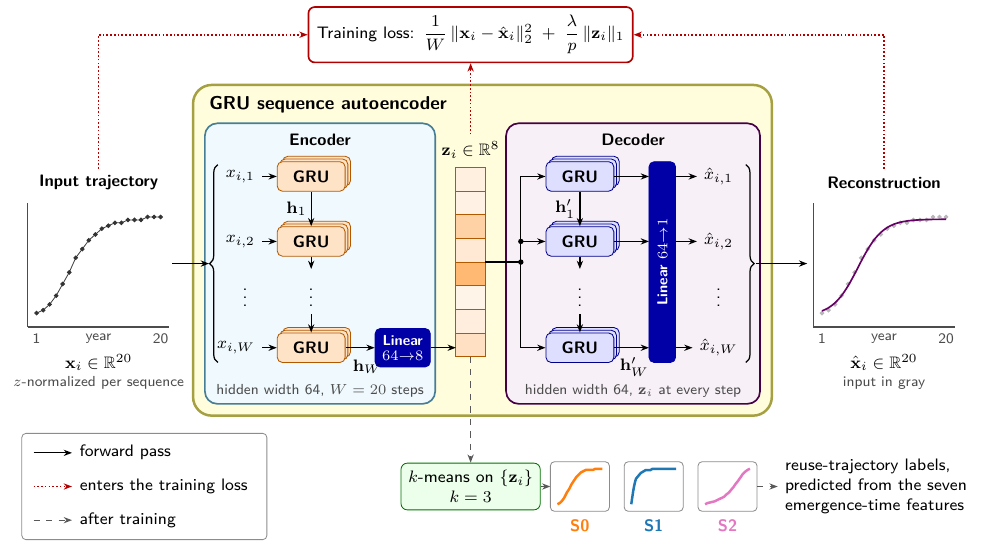}
\caption{Architecture of the GRU sequence autoencoder. The encoder, a single-layer GRU of hidden width $64$, reads the $z$-normalized trajectory $\mathbf{x}_i$ one year at a time, and a linear layer maps its final state $\mathbf{h}_W$ to the code $\mathbf{z}_i \in \mathbb{R}^{8}$, shaded by each coordinate's mean absolute value over the corpus. The decoder, a second single-layer GRU of width $64$, receives $\mathbf{z}_i$ at every step, and a linear layer shared across steps maps each state $\mathbf{h}'_t$ to $\hat{x}_{i,t}$. Training minimizes Equation~\eqref{eq:seqae}; afterwards, $k$-means on the codes yields the labels S0, S1 and S2, sketched by their median trajectories (Figure~\ref{fig:gru-shapes}).}
  \label{fig:gru-ae}
\end{figure}

Our primary labeling route derives reuse-trajectory labels from the trajectories themselves and uses none of the hand-crafted features. Shape-based methods for this purpose fix a distance in advance, typically an elastic one such as dynamic time warping \citep{berndt1994dtw, petitjean2011global}; we instead learn the representation from the raw sequences, for the reasons set out in Section~\ref{sec:related-seqclust}. Figure~\ref{fig:gru-ae} summarizes the architecture and the labeling step that follows it.

Each trajectory is $z$-normalized per sequence, so that clustering responds to shape rather than magnitude, giving $\mathbf{x}_i = (x_{i,1},\dots,x_{i,W})$. A gated recurrent unit (GRU) encoder \citep{cho2014gru} reads the sequence one step at a time, maintaining a hidden state $\mathbf{h}_t \in \mathbb{R}^{d_h}$ through an update gate $\mathbf{u}_t$ and a reset gate $\mathbf{r}_t$:
\begin{align}
\mathbf{u}_t &= \sigma\!\left(\mathbf{W}_u x_{i,t} + \mathbf{U}_u \mathbf{h}_{t-1} + \mathbf{b}_u\right), \label{eq:gru-update}\\
\mathbf{r}_t &= \sigma\!\left(\mathbf{W}_r x_{i,t} + \mathbf{U}_r \mathbf{h}_{t-1} + \mathbf{b}_r\right), \label{eq:gru-reset}\\
\tilde{\mathbf{h}}_t &= \tanh\!\left(\mathbf{W}_h x_{i,t} + \mathbf{U}_h \left(\mathbf{r}_t \odot \mathbf{h}_{t-1}\right) + \mathbf{b}_h\right), \label{eq:gru-cand}\\
\mathbf{h}_t &= \left(\mathbf{1} - \mathbf{u}_t\right) \odot \mathbf{h}_{t-1} + \mathbf{u}_t \odot \tilde{\mathbf{h}}_t, \label{eq:gru-state}
\end{align}
where $\sigma$ is the logistic function, $\odot$ is elementwise multiplication and $\mathbf{h}_0 = \mathbf{0}$. We write the update gate as $\mathbf{u}_t$ rather than the customary $\mathbf{z}_t$ to keep $\mathbf{z}$ for the latent code. The final state is projected to that code,
\begin{equation}
\label{eq:latent}
\mathbf{z}_i = \mathbf{W}_e \mathbf{h}_W + \mathbf{b}_e \in \mathbb{R}^{p}, \qquad p = 8 ,
\end{equation}
and a second GRU, conditioned on $\mathbf{z}_i$ at every step rather than at initialization alone, reconstructs the sequence as $\hat{x}_{i,t} = \mathbf{w}_o^{\top} \mathbf{h}'_t + b_o$ with $\mathbf{h}'_t = \mathrm{GRU}_{\mathrm{dec}}(\mathbf{z}_i, \mathbf{h}'_{t-1})$. Training minimizes reconstruction error with an $\ell_1$ penalty on the code,
\begin{equation}
\label{eq:seqae}
\mathcal{L}_{\mathrm{seq}} = \frac{1}{n}\sum_{i=1}^{n} \left[ \frac{1}{W} \left\| \mathbf{x}_i - \hat{\mathbf{x}}_i \right\|_2^2 + \frac{\lambda}{p} \left\| \mathbf{z}_i \right\|_1 \right], \qquad \lambda = 0.01 ,
\end{equation}
which encourages a sparse, cluster-friendly representation. We use a plain GRU rather than the orthogonally constrained variant of \citet{zadorozhnyy2024orthogonal} because our sequences are short ($W = 20$ steps), so long-range gradient behavior is not a concern. However, for problems with long sequences, implementing the orthogonal variant may be beneficial.

After training, we partition the codes by $k$-means,
\begin{equation}
\label{eq:kmeans-latent}
\min_{\{\mathcal{C}_c\},\,\{\boldsymbol{\mu}_c\}} \; \sum_{c=1}^{k} \sum_{i \in \mathcal{C}_c} \left\| \mathbf{z}_i - \boldsymbol{\mu}_c \right\|_2^2 ,
\end{equation}
and fix $k$ by the elbow criterion on this objective, with silhouette reported alongside and the alternative count examined directly (Section~\ref{sec:results-gru}). Because these labels are computed from the trajectories alone, predicting them from the seven emergence-time features is a prediction task with a genuinely external target, not a re-derivation of the labeling. Architecture and training details are given in Section~\ref{sec:setup}.

\subsection{Emergence-Profile Labeling via Feature Selection and Clustering}
\label{sec:method-fae}
\label{sec:method-clustering}
The second route labels technologies by their profile at emergence, in two steps: it selects a subset of the seven features and then clusters technologies on that subset. Not all seven features contribute equally to a meaningful partition of technologies, and redundant or noisy features can degrade both clustering and downstream classification. We therefore select an informative subset using the Fractal Autoencoder (FAE) \citep{wu2021fractal}, an unsupervised method that combines global reconstruction with a diversity-promoting sub-network.

FAE augments a linear autoencoder with a one-to-one scoring layer. Let $\mathbf{W}_I = \mathrm{Diag}(\mathbf{w})$, $\mathbf{w} \in \mathbb{R}^{m}$, be non-negative per-feature importance weights, and let $g(\mathbf{X}) = \mathbf{X}\mathbf{W}_E$ and $f(\mathbf{Z}) = \mathbf{Z}\mathbf{W}_D$ be a linear encoder and decoder with $\mathbf{W}_E \in \mathbb{R}^{m \times d}$, $\mathbf{W}_D \in \mathbb{R}^{d \times m}$, where $d$ is the encoder width. Writing $\mathbf{W}_I^{\max_K}$ for the operator that retains the $K$ largest entries of $\mathbf{W}_I$ and zeroes the rest, FAE minimizes
\begin{equation}
\label{eq:fae}
\mathcal{L} =
\underbrace{\bigl\| \mathbf{X} - f\!\left(g\!\left(\mathbf{X}\mathbf{W}_I\right)\right) \bigr\|_F^2}_{\text{global reconstruction}}
+ \lambda_1
\underbrace{\bigl\| \mathbf{X} - f\!\left(g\!\left(\mathbf{X}\mathbf{W}_I^{\max_K}\right)\right) \bigr\|_F^2}_{\text{sub-network reconstruction}}
+ \lambda_2 \|\mathbf{W}_I\|_1,
\quad \text{s.t. } \mathbf{W}_I \geq 0 .
\end{equation}
The first term learns globally representative weights; the second requires that the $K$ highest-weighted features alone reconstruct the data, promoting a diverse and non-redundant subset; the $\ell_1$ term encourages sparsity. After training, the $K$ features with the largest importance weights are selected. We use the linear variant with the original hyperparameters \citep{wu2021fractal} and evaluate $K \in \{3,4,5\}$. FAE is fit on the full corpus: feature selection is part of the unsupervised labeling stage, which precedes and is independent of the supervised train/test split (Section~\ref{sec:setup}).

Given the FAE-selected features, we standardize them to zero mean and unit variance, writing $\mathbf{s}_i \in \mathbb{R}^{K}$ for the standardized selected features of technology $i$, and partition technologies by $k$-means with $k$-means++ initialization,
\begin{equation}
\label{eq:kmeans-feat}
\min_{\{\mathcal{C}_c\},\,\{\boldsymbol{\nu}_c\}} \; \sum_{c=1}^{k} \sum_{i \in \mathcal{C}_c} \left\| \mathbf{s}_i - \boldsymbol{\nu}_c \right\|_2^2 .
\end{equation}
Equations~\eqref{eq:kmeans-latent} and~\eqref{eq:kmeans-feat} are the same objective over different spaces, and that is precisely the point of comparing them: the first partitions technologies by how their reuse unfolded, the second by what they looked like at emergence. We keep the standard algorithm rather than a variant that selects the number of clusters automatically \citep{sinaga2020unsupervised}, so that the cluster count is set by explicit, reported criteria. Each technology receives the label of its assigned cluster. To choose the number of clusters we sweep $k \in \{2,\dots,6\}$ under three internal validity criteria (Section~\ref{sec:results-clustering}, Table~\ref{tab:k-sweep}): the Davies--Bouldin index attains its optimum at $k=3$, and every partition beyond $k=3$ splits off clusters holding at most $1.1\%$ of the corpus. We therefore use $k=3$. Like FAE, the clustering is fit on the full corpus, and the train/test split is applied afterwards.

\subsection{Supervised Classification}
\label{sec:method-classification}
Given a labeling $y : \{1,\dots,n\} \to \{1,\dots,k\}$ from either route, we fit a classifier $\hat{y}_\theta$ that outputs a distribution over the $k$ classes by minimizing regularized cross-entropy on the training split,
\begin{equation}
\label{eq:erm}
\hat{\theta} = \arg\min_{\theta} \; -\frac{1}{|\mathcal{D}_{\mathrm{tr}}|} \sum_{i \in \mathcal{D}_{\mathrm{tr}}} \sum_{c=1}^{k} \mathbb{I}\!\left[y_i = c\right] \log p_\theta\!\left(c \mid \mathbf{x}_i\right) + \Omega(\theta) ,
\end{equation}
where $p_\theta(\cdot \mid \mathbf{x}_i)$ is the predicted class distribution and $\Omega$ collects whatever regularization the architecture applies. The architectures differ only in how $p_\theta$ is parameterized, so Equation~\eqref{eq:erm} is common to all seven. We evaluate two tree ensembles, Gradient Boosting Decision Trees (GBDT) and Extremely Randomized Trees (ExtraTrees), and five deep tabular models: TabNet \citep{arik2021tabnet}, FT-Transformer \citep{gorishniy2021revisiting}, TabM \citep{gorishniy2025tabm}, the Kolmogorov--Arnold network TabKAN \citep{eslamian2025tabkan}, and the MLP-mixer variant TabMixer \citep{eslamian2025tabmixer}. GBDT is the model reported best by \citet{chen2025predicting} and serves as our primary baseline. Because the classes are unbalanced, we report one-vs-rest macro ROC-AUC alongside accuracy and macro-$F_1$, and per-class figures where a minority class carries the interpretation. Training protocol and full hyperparameters are given in Section~\ref{sec:setup} and Appendix~\ref{app:hyperparams}.

\subsection{Validation Methodology}
\label{sec:method-validation}
Because the feature-based cluster labels are induced from the same features the classifier receives as input, a natural question is whether high classification accuracy reflects learned structure or merely the deterministic nature of the labeling. We address this with two complementary validations. The sequence-based labels of Section~\ref{sec:method-gru} do not share this concern, since they are derived from information the classifier never observes; for them, the classification test itself plays the role of the transfer test below.

\paragraph{Visual case study.} Following the practice of inspecting representative predictions in classification studies, we draw a random sample of held-out test technologies from each cluster and plot their reuse trajectories together with their assigned (clustering) and predicted (classifier) labels. Agreement between the trajectory shape, the assigned label, and the predicted label provides a direct, human-interpretable check that classifications are meaningful.

\paragraph{Independent labeling scheme.} We construct a second set of labels that does not depend on our features, by clustering the reuse \emph{trajectories} rather than the features, following the shape-based idea of \citet{chen2025predicting}. Concretely, we apply $k$-means to the $z$-normalized trajectories to obtain trajectory-shape labels for every technology, with $k{=}4$ to match \citet{chen2025predicting}. \citet{chen2025predicting} cluster trajectories with DTW-based $k$-means; we do not have access to their cluster labels, so we use Euclidean $k$-means on the $z$-normalized trajectories, which their appendix reports as agreeing with the DTW clustering on $92.85\%$ of technologies. Our labeling therefore reproduces their labeling procedure, up to the choice of distance, on our own corpus rather than their labels themselves. Because their score was obtained on a different patent corpus, period and labeling, we treat their published ROC-AUC as a reference point rather than a directly comparable result. The trajectory-shape labels are derived from information disjoint from the classifier's inputs: the classifier never observes trajectories. We use this independently constructed labeling in two ways.

\emph{Transfer without retraining.} We first take the classifier trained on the feature-based labels and, without retraining, compare its held-out predictions against the trajectory-shape labels. Because the two labelings need not use the same number of clusters, we quantify agreement with the Adjusted Rand Index (ARI) and Normalized Mutual Information (NMI), which compare partitions independently of label identity, and we additionally report ARI at matched cluster counts. Both outcomes of this test are informative. High agreement would mean that our feature-based partition is close to a trajectory-shape partition, so that the supervised stage has effectively recovered trajectory structure. Low agreement would mean the two partitions describe different structure, and therefore that the near-perfect accuracy of the supervised stage must not be read as a rediscovery of trajectory shape.

\emph{Direct evaluation on the trajectory task.} We then retrain each classifier to predict the trajectory-shape labels from the same emergence-time features, and report one-vs-rest macro ROC-AUC. \citet{chen2025predicting} report ROC-AUC without stating an averaging convention, so we fix ours explicitly and treat their score as a reference point rather than a like-for-like comparison. Because the trajectory labels are constructed without reference to the features, this is a prediction task with a genuinely external target. This evaluation, and not the accuracy on our own emergence-profile labels, is what makes the feature-based route's numbers interpretable; the paper's main empirical claim rests on the sequence-based task of Section~\ref{sec:method-gru}, whose target is likewise external to the features.

\section{Experimental Setup}
\label{sec:setup}

\subsection{Dataset}
We collected invention patents from the United States Patent and Trademark Office (USPTO) Open Data Portal for the period 2002--2022. Patents are a long-established if imperfect indicator of inventive activity \citep{griliches1990patent}. Following \citet{pezzoni2022fast} and \citet{chen2025predicting}, we define a \emph{technology} as a pairwise combination of six-digit International Patent Classification (IPC) codes, a \emph{novel technology} as the first appearance of such a pair, and \emph{early inventions} as the invention patents applying that technology within the first year of its emergence. New co-occurrences of classification codes on patents are also a common marker of technology fusion and convergence \citep{caviggioli2016technology, lee2022comparing}, which has likewise been traced through citation flows between patent classes \citep{no2010trajectory, karvonen2013patent}. Convergence studies usually require the codes to come from previously distinct fields; our novel technologies include such cross-field pairings but are not restricted to them. We construct each technology's reuse trajectory as the cumulative count of subsequent inventions over a 20-year observation window, and retain technologies with at least 20 reuses within that window, matching the filtering criterion of \citet{chen2025predicting}. Reuse is counted from classification codes rather than citations, following \citet{chen2025predicting} and \citet{pezzoni2022fast}. Forward citations are a validated indicator of technologically important patents \citep{albert1991direct}, but a large share of them are added by examiners \citep{alcacer2006patent}; code-based reuse captures every later patent that applies the same pairing whether or not it cites the originating one, at the cost of depending on classification practice instead. The filter yields $201{,}710$ novel technologies, each described by the seven features of Table~\ref{tab:features}.

\paragraph{Observation window and right-censoring.} Follow-on activity accrues over many years with a peaked age profile \citep{mehta2010identifying}, so a long observation window is needed. Emergence years in the corpus range from 2002 to 2021, so only technologies emerging by 2003 ($68{,}909$, or $34.2\%$) are observed for the full 20-year window; for later cohorts, years beyond the 2022 data horizon contribute no further reuse, and their cumulative trajectories are flat by construction from that point on. Two properties bound the impact of this censoring. First, the reuse filter itself concentrates the corpus in early cohorts: the 2002--2004 cohorts contain roughly half of all technologies, later cohorts shrink rapidly (the 2021 cohort contains only 67), and the median observed span is 19 of the 20 window years. Second, we quantify where the censored technologies end up in the labeling: the late-emerging, most heavily censored technologies concentrate in a single small cluster, which we characterize in Section~\ref{sec:results-clustering}. The reuse filter also means the corpus conditions on success: technologies that never reached 20 reuses are excluded. We return to the implications of the censoring and of this filter in Section~\ref{sec:discussion}.

\subsection{Features}
The seven features (Table~\ref{tab:features}) characterize a technology at its emergence year along two axes: the accessibility and similarity of its technology components, and the applicability and attention of its early inventions. Two of these constructs have established analogues: \texttt{SIM\_TECH} is a hierarchical IPC distance of the kind used to measure technological distance between patent classes \citep{yan2017measuring}, and \texttt{INVENT\_DIVER}, the spread of early inventions across IPC sections, corresponds to the pervasiveness criterion used to identify general purpose technologies \citep{bekar2018general, petralia2020mapping}. We compute them exactly as defined by \citet[Eqs.~3--9]{chen2025predicting}. We note that \texttt{SIM\_TECH} takes only four distinct values, a direct consequence of its hierarchical IPC-distance definition; this property is discussed in Section~\ref{sec:discussion}.

\begin{table}[htbp]
\centering
\caption{The seven technology-component and early-invention features (\citealp{chen2025predicting}).}
\label{tab:features}
\begin{tabular}{@{}lll@{}}
\toprule
Feature & Dimension & Description \\
\midrule
\texttt{ACCESS\_SIZE}  & Accessibility & Component patent counts, 5 yr pre-emergence \\
\texttt{ACCESS\_TREND} & Accessibility & Growth ratio of component accessibility \\
\texttt{SIM\_ACCESS}   & Similarity    & Difference in component cumulative counts \\
\texttt{SIM\_TECH}     & Similarity    & Weighted IPC hierarchy distance \\
\texttt{INVENT\_DIVER} & Applicability & Entropy of IPC sections in early inventions \\
\texttt{INVENT\_APPL}  & Applicability & Mean IPC codes per early invention \\
\texttt{ATTENT\_SIZE}  & Attention     & Number of early inventions \\
\bottomrule
\end{tabular}
\end{table}

\subsection{Pipeline and Protocol}
The emergence-profile pipeline has three stages: (i) unsupervised feature selection with the Fractal Autoencoder (FAE) \citep{wu2021fractal}; (ii) $k$-means clustering on the selected features to assign emergence-profile labels; and (iii) supervised classification of those labels from the same features. The sequence-based pipeline replaces stages (i) and (ii) with the recurrent autoencoder of Section~\ref{sec:method-gru}, applied to the trajectories, and keeps stage (iii) unchanged. Stages (i) and (ii) are unsupervised and are fit on the full corpus; they constitute the label construction, and no supervised model is involved. The train/test split is applied afterwards, to the classification stage only: we use a stratified 60/40 train/test split, with a further 90/10 train/validation split within the training set for early stopping. The comparatively large test fraction gives tight test-set estimates and follows the protocol under which all our supervised results, including the trajectory-task comparisons of Section~\ref{sec:results-transfer}, were produced. All splits use a fixed random seed. Classifier input features are standardized using statistics computed on the training set only.

\subsection{FAE Configuration}
We use the linear variant of FAE, minimizing the objective of Eq.~\ref{eq:fae} with the original hyperparameters \citep{wu2021fractal}: encoder width equal to the selection budget ($d = K$), $\lambda_1 = 2$, $\lambda_2 = 0.1$, Adam optimizer with learning rate $10^{-3}$, $3000$ training epochs, and feature-importance weights initialized from $\mathcal{U}[0.999999, 0.9999999]$. A $1000$-epoch run selects identical features (Section~\ref{sec:results-fae}).

\subsection{GRU Autoencoder Configuration}
The sequence autoencoder uses a single-layer GRU encoder with hidden width $64$ mapped linearly to a latent of dimension $8$, and a single-layer GRU decoder conditioned on the latent at every step. It is trained on the per-sequence $z$-normalized trajectories with the Adam optimizer (learning rate $10^{-3}$), batch size $512$, $40$ epochs, latent $\ell_1$ weight $\lambda = 0.01$, and the same fixed random seed as the rest of the pipeline. The configuration was selected from a grid over $\{1, 3\}$ encoder layers, latent dimension $\{8, 16, 32\}$, and $\ell_1$ weight $\{0, 0.01\}$: reconstruction error is low throughout (mean squared error $0.001$--$0.004$ on the normalized sequences), deeper encoders reduce reconstruction error without improving cluster separation, and the $\ell_1$ penalty raises the latent-space silhouette of the resulting clustering from $0.35$ to $0.47$ at this configuration. Because cluster separation informed this model selection, we treat the elbow on the $k$-means cost, not silhouette, as the primary criterion for the cluster count (Section~\ref{sec:results-gru}). Like the feature-based route, the autoencoder and the $k$-means step are fit on the full corpus as part of label construction; the supervised train/test split is applied afterwards. For classification of the sequence-based labels we use the same stratified 60/40 split and training-set standardization as elsewhere; the tree ensembles are fit on the full training portion, as no early-stopping validation split is needed.

\subsection{Classifiers}
We evaluate seven tabular classifiers spanning tree ensembles and deep tabular architectures: Gradient Boosting Decision Trees (GBDT) and Extremely Randomized Trees (ExtraTrees); TabNet \citep{arik2021tabnet}; FT-Transformer \citep{gorishniy2021revisiting}; TabM \citep{gorishniy2025tabm}; the Kolmogorov--Arnold network TabKAN \citep{eslamian2025tabkan}; and the MLP-mixer variant TabMixer \citep{eslamian2025tabmixer}. GBDT is included as the model reported best by \citet{chen2025predicting}.

\subsection{Metrics}
We report classification accuracy and macro-averaged $F_1$ (macro-$F_1$) and, where relevant, ROC-AUC, the metric for which \citet{chen2025predicting} publish a score that we use as a reference point. Clustering quality is assessed with three complementary metrics: the silhouette coefficient \citep{rousseeuw1987silhouette}, the Davies--Bouldin index \citep{davies1979bouldin} (lower is better), and the Calinski--Harabasz index \citep{calinski1974harabasz} (higher is better).

\section{Results}
\label{sec:results}
We report the emergence-profile route first, not because it carries the main claim but because it supplies the comparison the main claim is measured against. Sections~\ref{sec:results-fae} to~\ref{sec:results-casestudy} characterize that route and establish that a classifier recovers its labels almost perfectly, which is a consistency check rather than a forecast, since the labels and the classifier share the same seven features. Section~\ref{sec:results-transfer} then asks what those labels actually describe by comparing them with a labeling built from the trajectories, and Section~\ref{sec:results-gru} presents the sequence-based route, whose labels are constructed without reference to the features and whose recovery from them is therefore a genuine prediction.

\subsection{Feature Selection}
\label{sec:results-fae}

\begin{table}[htbp]
\centering
\caption{FAE-selected features and reconstruction losses for each budget $K$.}
\label{tab:fae-selection}
\begin{tabular}{@{}clr@{}}
\toprule
$K$ & Selected features & Total loss \\
\midrule
3 & \texttt{SIM\_TECH}, \texttt{ACCESS\_SIZE}, \texttt{SIM\_ACCESS} & $2{,}018{,}093$ \\
4 & \texttt{ACCESS\_TREND}, \texttt{SIM\_ACCESS}, \texttt{INVENT\_DIVER}, \texttt{INVENT\_APPL} & $1{,}039{,}342$ \\
5 & \texttt{ACCESS\_SIZE}, \texttt{INVENT\_APPL}, \texttt{ACCESS\_TREND}, \texttt{SIM\_TECH}, \texttt{INVENT\_DIVER} & $633{,}309$ \\
\bottomrule
\end{tabular}
\end{table}

We first apply FAE to the seven emergence-time features and record which features it selects for each budget $K \in \{3,4,5\}$, along with the reconstruction loss at each budget (Table~\ref{tab:fae-selection}). The selected subsets are not nested: the $K{=}4$ selection does not simply add a feature to the $K{=}3$ selection, because FAE re-optimizes the importance weights independently for each $K$. The reconstruction loss decreases as $K$ grows, since more features are available to reconstruct the original data. The selection is stable to training length: FAE run for $1000$ and for $3000$ epochs produces identical selections. Robustness to data resampling is a stricter criterion, which we quantify in Section~\ref{sec:discussion}. The $K{=}3$ subset consists entirely of technology-component features (\texttt{SIM\_TECH}, \texttt{ACCESS\_SIZE}, \texttt{SIM\_ACCESS}); all three early-invention features are omitted, a point that becomes important for the trajectory task in Section~\ref{sec:results-transfer}.

\subsection{Clustering Quality}
\label{sec:results-clustering}

\paragraph{Choosing the number of clusters.}

\begin{table}[htbp]
\centering
\caption{Cluster-count sweep on the standardized FAE $K{=}3$ features. Silhouette is computed on a $10{,}000$-point sample. The selected $k{=}3$ is in bold.}
\label{tab:k-sweep}
\begin{tabular}{@{}crrrr@{}}
\toprule
$k$ & Silhouette ($\uparrow$) & Calinski--Harabasz ($\uparrow$) & Davies--Bouldin ($\downarrow$) & Smallest cluster \\
\midrule
2 & $0.568$ & $92{,}529$  & $0.810$ & $43.8\%$ \\
\textbf{3} & $\mathbf{0.646}$ & $\mathbf{244{,}583}$ & $\mathbf{0.544}$ & $\mathbf{3.4\%}$ \\
4 & $0.674$ & $289{,}709$ & $0.580$ & $1.1\%$ \\
5 & $0.681$ & $302{,}713$ & $0.605$ & $0.3\%$ \\
6 & $0.685$ & $341{,}246$ & $0.586$ & $0.3\%$ \\
\bottomrule
\end{tabular}
\end{table}

We cluster the standardized FAE $K{=}3$ features with $k$-means, sweeping $k \in \{2,\dots,6\}$ under three internal validity criteria (Table~\ref{tab:k-sweep}). The Davies--Bouldin index attains its optimum at $k{=}3$. Silhouette and Calinski--Harabasz increase monotonically across the swept range and therefore do not identify an interior optimum, but the marginal silhouette gain falls sharply after $k{=}3$ ($+0.078$ from $k{=}2$ to $3$ versus $+0.029$ from $3$ to $4$), and every partition beyond $k{=}3$ splits off clusters holding at most $1.1\%$ of the corpus. We therefore use $k{=}3$: the best partition under Davies--Bouldin, and the largest $k$ at which every cluster still holds a substantial share of the corpus.

\paragraph{The selected subset against all alternatives.}

\begin{table}[htbp]
\centering
\caption{Rank of the FAE $K{=}3$ feature subset among all 35 three-feature subsets, by clustering metric ($k{=}3$). Lower rank is better.}
\label{tab:cluster-metrics}
\begin{tabular}{@{}lrr@{}}
\toprule
Metric & FAE value & Rank (of 35) \\
\midrule
Silhouette ($\uparrow$)        & $0.646$       & 11 \\
Calinski--Harabasz ($\uparrow$) & $244{,}583$   & 1 \\
Davies--Bouldin ($\downarrow$)  & $0.544$       & 2 \\
\bottomrule
\end{tabular}
\end{table}

Because the silhouette coefficient is only one of several ways to measure clustering quality, we compare the FAE-selected subset against all $\binom{7}{3}=35$ possible three-feature subsets under three separate metrics at $k{=}3$ (Table~\ref{tab:cluster-metrics}). The FAE subset ranks first of $35$ under Calinski--Harabasz, second under Davies--Bouldin, and eleventh under silhouette: by the variance-based criteria, the selected features produce well-separated clusters. The divergence between silhouette and the variance-based criteria is analyzed in Section~\ref{sec:discussion}.

\begin{figure}[htbp]
  \centering
  \includegraphics[width=0.8\linewidth]{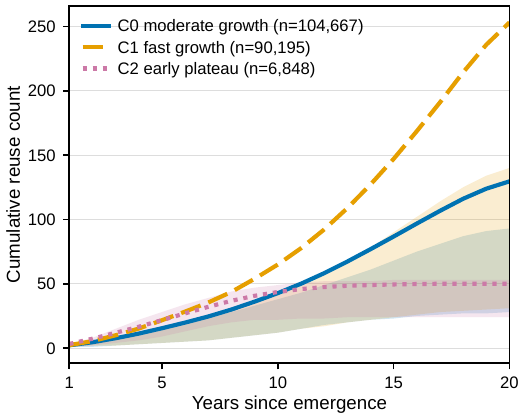}
\caption{Mean cumulative reuse trajectory of each cluster (FAE $K{=}3$, $k{=}3$); shaded bands show the 25--75 percentile range.}
  \label{fig:trajectories}
\end{figure}

\paragraph{What the clusters are.}

\begin{table}[htbp]
\centering
\caption{Composition of the three clusters (FAE $K{=}3$, $k{=}3$). Mean reuse is the cluster mean at year 20, reported both over the full corpus and over the 2005 and later cohorts, the range in which all three clusters occur: cluster 2 has no members at all in the 2002--2004 cohorts, so the full-corpus column compares populations of different ages. The last two columns give the median emergence year and the two most common IPC sections, with the percentage share of the cluster's component codes in parentheses.}
\label{tab:cluster-composition}
\setlength{\tabcolsep}{3pt} \footnotesize
\begin{tabular}{@{}clrrrcc@{}}
\toprule
Cluster & Pattern & Size & \multicolumn{2}{c}{Mean reuse at yr 20} & Median yr & Top IPC sections \\
\cmidrule(lr){4-5}
 & & & all & 2005+ & & \\
\midrule
0 & Moderate growth & $104{,}667$ ($51.9\%$) & $130$ & $56$ & 2005 & B ($21.4$), G ($15.5$) \\
1 & Fast growth      & $90{,}195$  ($44.7\%$) & $253$ & $65$ & 2004 & B ($23.8$), C ($19.7$) \\
2 & Early plateau    & $6{,}848$   ($3.4\%$)  & $50$  & $50$ & 2014 & G ($33.3$), H ($28.3$) \\
\bottomrule
\end{tabular}
\end{table}

The three clusters differ in their \emph{mean} reuse trajectories (Table~\ref{tab:cluster-composition}, Figure~\ref{fig:trajectories}): a large moderate-growth group, a comparably large fast-growth group, and a small early-plateau group. We name them by those averages for readability only, and two cautions attach to the names. Section~\ref{sec:results-transfer} shows the separation does not survive at the level of individual technologies, so the names describe emergence-time profiles and not a reuse taxonomy. The averages themselves are also inflated by cohort age: cluster 2 has no members in the 2002--2004 cohorts, so over the full corpus it is compared against clusters observed for far longer. Within the 2005 and later cohorts, where all three occur, the mean year-20 totals are $56$, $65$ and $50$ rather than $130$, $253$ and $50$ (Table~\ref{tab:cluster-composition}), and what looked like a fivefold gap between the fast-growth and early-plateau groups is closer to a third.

Beyond their trajectories, the clusters differ in composition. The two large clusters draw on similar technology areas, their most common IPC sections being B and G, and B and C respectively, and they emerge early (median years 2005 and 2004). The small early-plateau cluster is different in kind: its technologies emerge late (median year 2014, with $65\%$ emerging in 2013 or later), concentrate in computing and electronics (IPC sections G and H together account for $62\%$ of its component codes, against $27.6\%$ and $29.9\%$ in the other clusters), and are built from very large component areas (mean \texttt{ACCESS\_SIZE} of roughly $16{,}600$ component patents, at least fourteen times the level of either other cluster).

That last multiple is inflated by the start of the data. \texttt{ACCESS\_SIZE} counts component patents in the five years before emergence, which for the earliest cohorts reaches back before 2002 and is therefore undercounted, and the two large clusters are predominantly early (Section~\ref{sec:limitations}). Restricting the comparison to technologies emerging in 2007 or later, where the five-year window lies wholly inside the data, the cluster means become $16{,}900$ against $2{,}700$ and $1{,}700$, a factor of six to ten rather than the fourteen to twenty-nine of the full corpus. The difference is real but substantially smaller than the uncorrected figure suggests.

Because most of its members are right-censored (Section~\ref{sec:setup}), part of this cluster's flat late-window trajectory reflects the observation horizon rather than a saturation of reuse. We therefore read it as \emph{recent technologies recombining mature, heavily patented computing and electronics components}, whose early reuse is rapid but whose long-run trajectory is still partly unobserved.

\subsection{Classification}
\label{sec:results-classification}

\begin{table}[htbp]
\centering
\caption{Test-set classification on FAE $K{=}3$ features (three clusters).}
\label{tab:classification}
\begin{tabular}{@{}lrr@{}}
\toprule
Model & Accuracy (\%) & Macro-$F_1$ (\%) \\
\midrule
TabNet         & 99.89 & 99.45 \\
GBDT           & 99.96 & 99.78 \\
ExtraTrees     & 99.98 & 99.91 \\
TabM           & 99.985 & 99.93 \\
FT-Transformer & 99.989 & 99.95 \\
TabKAN         & 99.989 & 99.95 \\
TabMixer       & 99.994 & \textbf{99.97} \\
\bottomrule
\end{tabular}
\end{table}

We train all seven classifiers to predict the cluster labels from the FAE $K{=}3$ features (Table~\ref{tab:classification}). Every model exceeds $99.4\%$ macro-$F_1$. With GBDT, the classifier reported best by \citet{chen2025predicting}, the three FAE-selected features attain $99.96\%$ accuracy and $99.78\%$ macro-$F_1$, against $99.85\%$ and $99.65\%$ with all seven features, so the selection costs nothing on this task. The high macro-$F_1$ is not carried by the majority classes alone: the smallest cluster ($3.4\%$ of technologies) is recovered nearly perfectly as well (TabMixer reaches $99.85\%$ recall on it, $2{,}735$ of $2{,}739$ test members). Differences between the seven models are within a few tenths of a percentage point, which reflects how separable the task is in the selected feature space; Section~\ref{sec:discussion} explains why near-saturation is the expected outcome of this evaluation.

\subsection{Sample Efficiency}
\label{sec:results-sample}

\begin{figure}[htbp]
  \centering
  \includegraphics[width=0.8\linewidth]{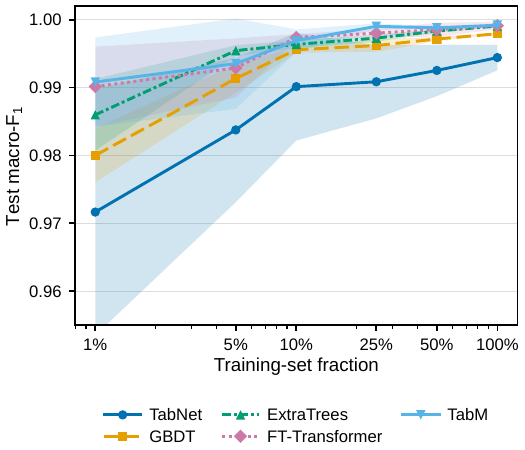}
\caption{Test macro-$F_1$ as the training-set fraction varies from $1\%$ to $100\%$ (mean over three seeds; shaded bands show one standard deviation). GBDT degrades faster than ExtraTrees, TabM, and FT-Transformer as data becomes scarce, and TabNet falls furthest; the differences are only pronounced at the smallest fractions. Exact values are given in Table~\ref{tab:fraction} in the appendix.}
  \label{fig:fraction}
\end{figure}

To probe behavior under limited supervision, we vary the training-set size from $1\%$ to $100\%$ while keeping the test set fixed, averaging over three seeds (Figure~\ref{fig:fraction}). We report this sweep for a representative subset of five classifiers, the two tree ensembles (GBDT and ExtraTrees) and three deep models (TabNet, TabM, and FT-Transformer); the two we omit here, TabKAN and TabMixer, sit within the same narrow band as the others at full data (Table~\ref{tab:classification}). From $5\%$ of the training data upward, all models except TabNet sit within a few tenths of a point of one another, with TabNet trailing by roughly a point. At $1\%$ ($1{,}089$ examples) the separation widens: GBDT drops to $98.0\%$ macro-$F_1$, below TabM ($99.1\%$), FT-Transformer ($99.0\%$), and the tree-based ExtraTrees ($98.6\%$), while TabNet falls furthest ($97.2\%$). The GBDT gap is architecture-specific rather than a deep-versus-tree effect, since ExtraTrees is itself a tree ensemble; Section~\ref{sec:discussion} discusses this and the TabNet outlier.

\subsection{Qualitative Case Study}
\label{sec:results-casestudy}

\begin{figure}[htbp]
  \centering
  \includegraphics[width=\linewidth]{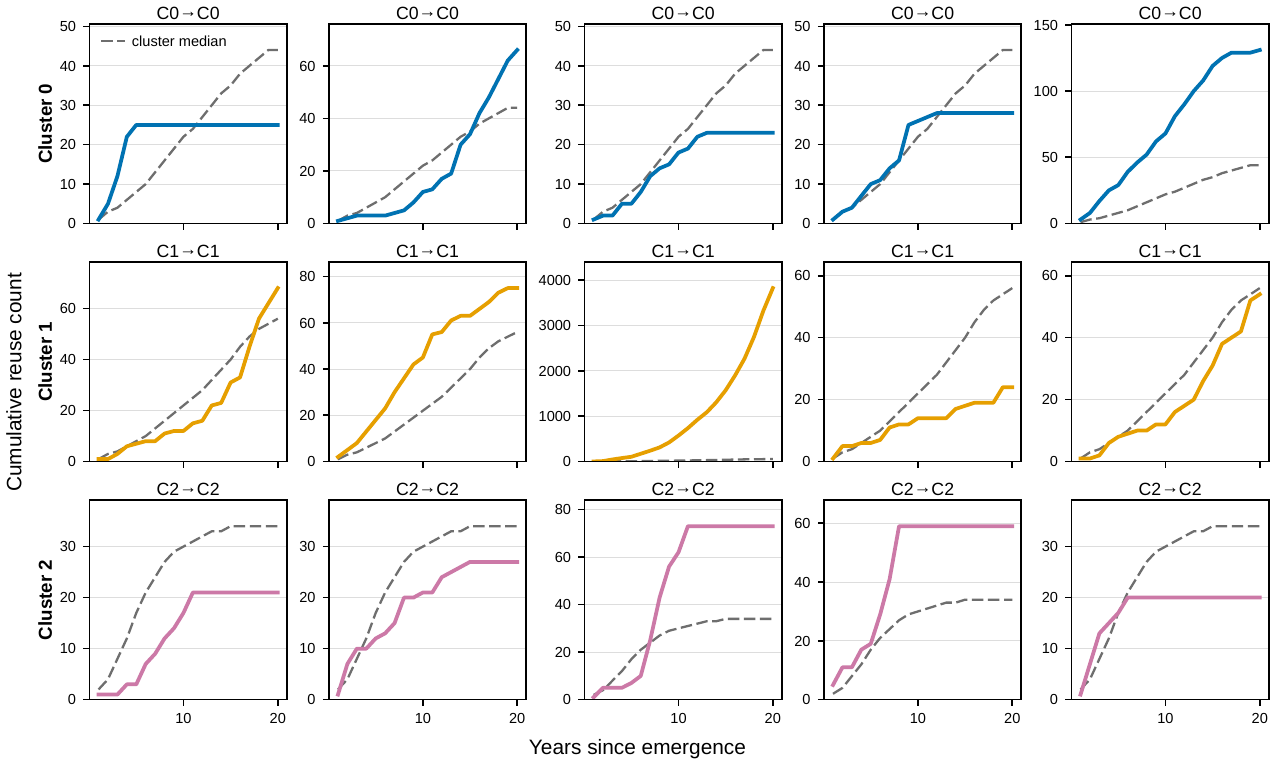}
\caption{Held-out test technologies sampled from each cluster (rows), with their assigned (clustering) and predicted (TabMixer) labels. Trajectories show cumulative reuse over the 20-year window; the dashed line is the cluster median. Panel $y$-axes are scaled individually so the shape of each trajectory is visible. All sampled examples are correctly classified.}
  \label{fig:casestudy}
\end{figure}

Before the quantitative validation, we inspect what the clusters contain. We randomly sample held-out test technologies from each cluster and plot their reuse trajectories, labeled with both their assigned (clustering) and predicted (classifier) cluster (Figure~\ref{fig:casestudy}). The sampled trajectories look like their cluster: the moderate-growth cluster accumulates steadily, the fast-growth cluster starts slowly and then accelerates to much higher totals, and the early-plateau cluster rises quickly before flattening out. The clustering only ever saw the emergence-time features, not the trajectories, yet the resulting groups correspond to visibly distinct reuse behaviors. The predicted labels agree with the assigned labels for every sampled example, consistent with the near-perfect test accuracy reported above.

Across the full test set of $80{,}684$ technologies, TabMixer misclassifies only $5$ ($0.006\%$). All five sit near the boundary between the early-plateau cluster and the moderate- or fast-growth clusters (Figure~\ref{fig:casestudy-err} in the appendix): they are technologies whose trajectories are atypical for their assigned cluster, such as an early-plateau technology that keeps accumulating steadily.

\subsection{Our Emergence-Profile Labeling Against Our Trajectory-Shape Labeling}
\label{sec:results-transfer}
Because our cluster labels and our classifier are built from the same seven features, the high accuracy in Table~\ref{tab:classification} could be a consequence of the labeling procedure rather than evidence of genuinely predictable structure. We address this in two steps: we first measure how far our labeling agrees with an externally constructed one, then we benchmark our models directly on the trajectory-prediction task.

\paragraph{Our emergence-profile labeling and our trajectory-shape labeling agree only marginally above chance.} We cluster the reuse \emph{trajectories} directly with $k$-means on the $z$-normalized curves (Section~\ref{sec:method-validation}) and measure how far that labeling agrees with our emergence-profile clustering. We quantify agreement with the Adjusted Rand Index \citep{hubert1985comparing} and Normalized Mutual Information \citep{vinh2009information}, which compare two partitions without regard to how their clusters are numbered. ARI is corrected for chance, so $0$ is the value expected of two independent partitions and $1$ means identical ones; NMI is normalized to $[0,1]$ with $1$ again meaning identical, but it is not chance-corrected, so independent partitions score above $0$ by an amount that grows with the number of clusters \citep{vinh2009information}. The chance-corrected figure is therefore the one to read.

At matched cluster counts the agreement is very low: ARI is $0.034$ for three emergence-profile clusters against three trajectory clusters, and $0.041$ for four against four. Comparing our three clusters against the four used by \citet{chen2025predicting} gives $0.028$ (NMI $0.044$), and the classifier's held-out predictions give essentially the same values as the cluster labels themselves ($\text{ARI}=0.028$, $\text{NMI}=0.045$), confirming that the classifier reproduces our labels rather than the trajectory labeling. Because the matched and mismatched comparisons agree, the low figure is not an artifact of comparing partitions of different sizes. The contingency table adds one nuance: the two large emergence-profile clusters spread across all four trajectory clusters in close to marginal proportions, while the small early-plateau cluster is genuinely concentrated, with $91\%$ of its members in the two earliest-saturating trajectory clusters, consistent with its censoring-heavy composition (Section~\ref{sec:results-clustering}).

This result is not in tension with Figure~\ref{fig:trajectories}, which compares cluster \emph{averages}: the average trajectories of the feature-based clusters are clearly separated (Figure~\ref{fig:trajectories}), while \emph{individual} assignments still disagree across the two partitions. The feature-based partition and the trajectory-shape partition capture largely different structure; the emergence-time features do not simply encode trajectory shape.

\paragraph{Direct evaluation on the trajectory task.}

\begin{figure}[htbp]
  \centering
  \includegraphics[width=0.72\linewidth]{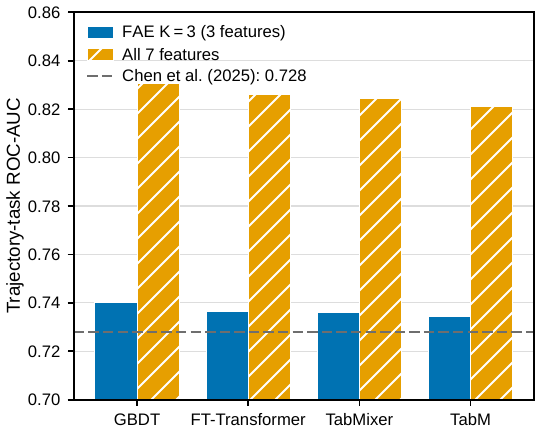}
\caption{Trajectory-task performance (one-vs-rest macro ROC-AUC, $k{=}4$ trajectory labels) for four classifiers retrained on the FAE $K{=}3$ features versus all seven features. The dashed line marks the reference GBDT score of $0.728$ from \citet{chen2025predicting}; all configurations fall above it, though that score was obtained on a different corpus and labeling and therefore serves as a reference point rather than a benchmark.}
  \label{fig:fae-vs-all7}
\end{figure}

Treating the trajectory labeling as an external target, we retrain each model to predict the trajectory-shape labels ($k{=}4$) from the emergence-time features and report one-vs-rest macro ROC-AUC, the metric reported by \citet{chen2025predicting}, whose published GBDT score of $0.728$ serves as a reference point (Section~\ref{sec:method-validation}). With the three FAE-selected features, GBDT reaches $0.740$, FT-Transformer $0.736$, TabMixer $0.736$, and TabM $0.734$, comparable to the reference. With all seven features, the same models reach $0.821$ to $0.831$ (GBDT $0.831$; Figure~\ref{fig:fae-vs-all7}), above the reference value, which was obtained on a different corpus and labeling (Section~\ref{sec:method-validation}).

\begin{table}[htbp]
\centering
\caption{Controlled ablation of \texttt{INVENT\_APPL} on the trajectory task (GBDT, one-vs-rest macro ROC-AUC, $k{=}4$ trajectory labels).}
\label{tab:ablation}
\begin{tabular}{@{}lcr@{}}
\toprule
Feature set & \#\,features & ROC-AUC \\
\midrule
FAE $K{=}3$                                   & 3 & $0.740$ \\
FAE $K{=}3$ $+$ \texttt{INVENT\_APPL}         & 4 & $0.768$ \\
All seven $-$ \texttt{INVENT\_APPL}           & 6 & $0.824$ \\
All seven                                     & 7 & $0.831$ \\
\bottomrule
\end{tabular}
\end{table}

\paragraph{A controlled ablation of the feature gap.}

\begin{figure}[htbp]
  \centering
  \includegraphics[width=0.72\linewidth]{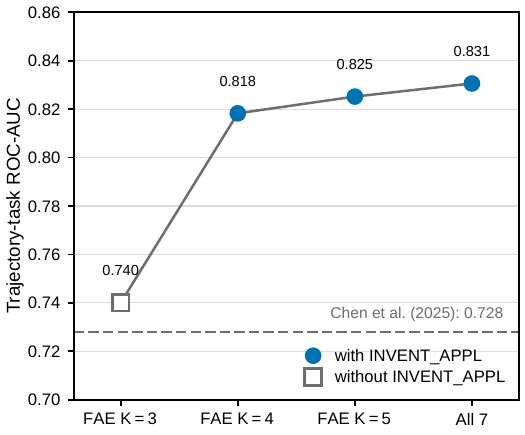}
\caption{GBDT trajectory-task ROC-AUC as a function of the FAE feature budget. Budgets that include \texttt{INVENT\_APPL} (filled circles) cluster near $0.82$; the FAE $K{=}3$ subset (open square), which omits it, sits at $0.740$, just above the reference score of $0.728$ from \citet{chen2025predicting}, which was obtained on a different corpus and labeling and therefore serves as a reference point rather than a benchmark. The controlled ablation in Table~\ref{tab:ablation} shows this association is not attributable to \texttt{INVENT\_APPL} alone.}
  \label{fig:feature-budget}
\end{figure}

Figure~\ref{fig:feature-budget} sweeps the GBDT feature budget: every budget containing \texttt{INVENT\_APPL} (FAE $K{=}4$, $K{=}5$, and all seven features) scores between $0.818$ and $0.831$, while the $K{=}3$ subset, the only one without it, reaches $0.740$. That association, however, is confounded, because the larger budgets add several features at once. We therefore run the controlled comparison directly (Table~\ref{tab:ablation}). Adding \texttt{INVENT\_APPL} alone to the $K{=}3$ subset, the single addition favored by the SHAP ranking of \citet{chen2025predicting}, raises the score from $0.740$ to $0.768$, about a third of the full gap. Removing \texttt{INVENT\_APPL} from the full set barely costs anything: $0.831$ to $0.824$. The seven-feature advantage is therefore \emph{distributed and redundant}: the information \texttt{INVENT\_APPL} carries is valuable but largely shared with the other early-invention features, and no single feature accounts for the gap. The deeper pattern is dimensional: the FAE $K{=}3$ subset contains only technology-component features, and what the trajectory task rewards is the early-invention information as a group.

\paragraph{The same cohort control on the trajectory task.} Section~\ref{sec:results-gru} shows that emergence year alone recovers much of the sequence-based labeling. The same control belongs here, because the trajectory-shape labels are built from the same right-censored curves. Holding the protocol fixed and varying only the predictors, emergence year on its own reaches a one-vs-rest macro ROC-AUC of $0.819$ against $0.831$ for the seven features, and on this task it exceeds them on the other two measures, at $63.3\%$ accuracy against $60.6\%$ and macro-$F_1$ $0.645$ against $0.591$. Dropping the two left-truncated features costs more here than on the sequence task, taking ROC-AUC to $0.761$. The cohort qualification therefore applies to both external tasks and not only to the sequence-based one, and we state it for both rather than letting the $0.831$ stand unqualified. It does not transfer to the reference score of \citet{chen2025predicting}, which was obtained on a different corpus and period whose censoring structure we have not examined, and which we continue to treat as a reference point rather than a comparison.

\subsection{Sequence-Based Labeling Results}
\label{sec:results-gru}

\begin{table}[htbp]
\centering
\caption{Sequence-based labeling as the cluster count varies. Silhouette is computed on the GRU latent (10{,}000-point sample); ROC-AUC is for GBDT predicting the sequence-based labels from all seven emergence-time features (one-vs-rest macro for $k \geq 3$, binary for $k{=}2$). The sweep ran over $k \in \{2,\dots,10\}$; $k = 2$ to $5$ are shown here and the full range appears in Figure~\ref{fig:gru-elbow}. The selected $k{=}3$ is in bold.}
\label{tab:gru-sweep}
\begin{tabular}{@{}crrr@{}}
\toprule
$k$ & Silhouette ($\uparrow$) & GBDT ROC-AUC & Smallest cluster \\
\midrule
2 & $0.512$ & $0.904$ & $23.4\%$ \\
\textbf{3} & $\mathbf{0.475}$ & $\mathbf{0.914}$ & $\mathbf{2.9\%}$ \\
4 & $0.449$ & $0.912$ & $0.4\%$ \\
5 & $0.326$ & $0.872$ & $0.2\%$ \\
\bottomrule
\end{tabular}
\end{table}

\paragraph{Choosing the number of clusters.}

\begin{figure}[htbp]
  \centering
  \includegraphics[width=0.95\linewidth]{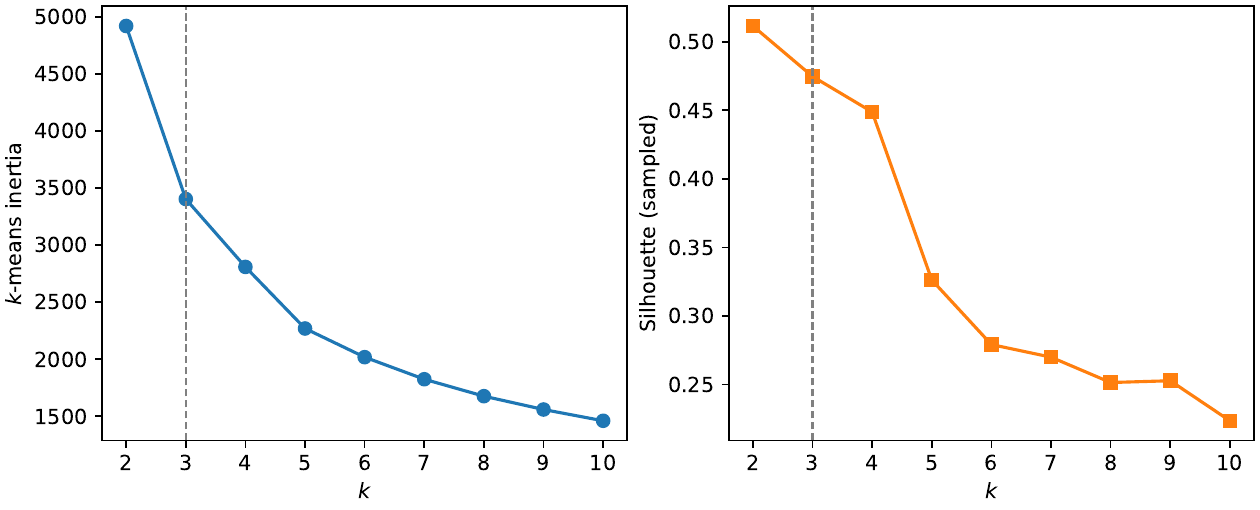}
\caption{Cluster-count selection for the sequence-based labeling: $k$-means cost on the GRU latent (left; the elbow at $k{=}3$ is marked) and sampled silhouette (right).}
  \label{fig:gru-elbow}
\end{figure}

We cluster the GRU latent vectors with $k$-means for $k \in \{2,\dots,10\}$ and apply the elbow criterion, our primary selection rule (Section~\ref{sec:setup}), to the $k$-means cost (Figure~\ref{fig:gru-elbow}). The cost curve has its elbow at $k{=}3$: the drop from $k{=}2$ to $3$ ($1{,}517$) is more than twice the drop from $3$ to $4$ ($595$). Silhouette does not by itself select $k{=}3$: it is highest at $k{=}2$ ($0.51$) and declines only gently through $k{=}3$ and $k{=}4$ ($0.47$, $0.45$) before falling sharply at $k{=}5$ ($0.33$).

We retain $k{=}3$ on grounds internal to the trajectories. The two-cluster partition absorbs nearly all of the early-plateau group into the mid-saturating cluster, and those two groups reach almost the same total reuse by year 20 (median $29$ against $32$) by very different routes, half of it within $2.7$ years against $6.9$ years. Collapsing them discards a timing distinction that is visible in the raw curves and is exactly the kind of structure a trajectory labeling exists to record. We do not appeal to the emergence-profile route here, since treating the two labelings as independently constructed forbids using either to justify the other.

To show that the downstream conclusion does not depend on this choice, we carry $k{=}2$ through the prediction stage as well: GBDT on all seven features recovers the two-cluster labeling at a binary ROC-AUC of $0.904$, against $0.914$ one-vs-rest macro at $k{=}3$ (Table~\ref{tab:gru-sweep}). The two figures are not the same quantity, since the task is binary in one case and three-class in the other, so we read them only as showing that emergence-time features anticipate the sequence-based partition at either count. The reference labeling of \citet{chen2025predicting} uses four trajectory clusters, likewise chosen by an elbow criterion, on DTW $k$-means over the raw trajectories; on the GRU latent the same criterion selects three.

\paragraph{What the sequence clusters are.}

\begin{figure}[htbp]
  \centering
  \includegraphics[width=0.8\linewidth]{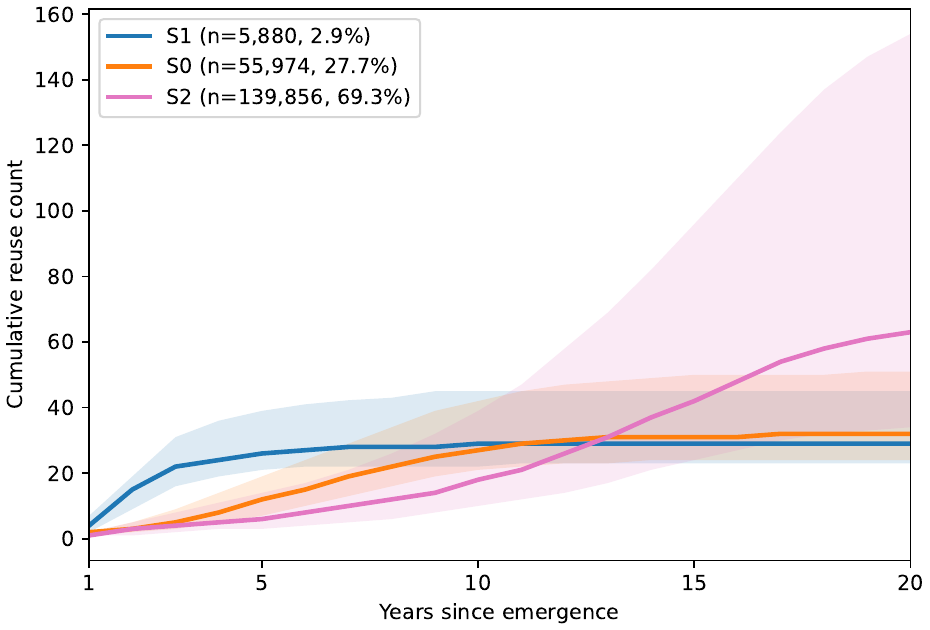}
\caption{Median reuse trajectory of each sequence-based cluster ($k{=}3$), with 25--75 percentile bands, on the raw (unnormalized) counts.}
  \label{fig:gru-shapes}
\end{figure}

The three clusters correspond to distinct temporal reuse patterns (Figure~\ref{fig:gru-shapes}). The largest, S2 ($139{,}856$ technologies, $69.3\%$), shows sustained growth that accelerates late in the window, reaching a median of $63$ reuses by year 20 and taking on average $13.4$ years to accumulate half of its final total; its members emerge early (median year 2004). S0 ($55{,}974$, $27.7\%$) grows moderately and saturates mid-window (half of its total by year $6.9$ on average). S1 ($5{,}880$, $2.9\%$) plateaus almost immediately (half of its total by year $2.7$), and $79.3\%$ of its members emerged in 2013 or later (median year 2017): like the early-plateau cluster of the feature-based route, it is dominated by right-censored recent technologies (Section~\ref{sec:setup}). That both labeling routes, built from disjoint information, isolate a small censoring-shaped cluster is itself a consistency check on the corpus.

\paragraph{The $k{=}4$ minority cluster.}

\begin{figure}[htbp]
  \centering
  \includegraphics[width=0.75\linewidth]{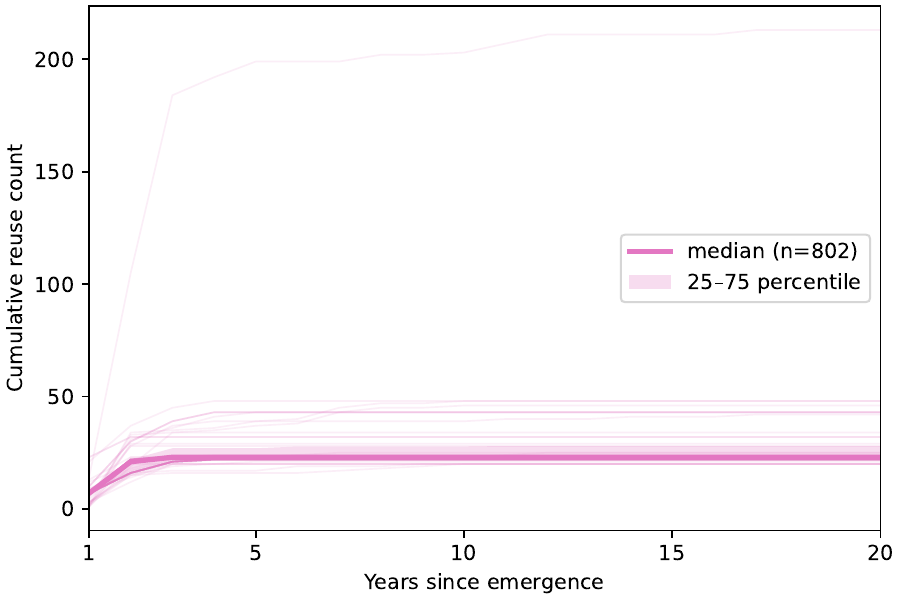}
\caption{The minority cluster that appears at $k{=}4$ ($802$ technologies, $0.40\%$): 60 randomly drawn member trajectories (light), median (bold), and 25--75 percentile band. The shape is consistent, and its members are overwhelmingly recent (median emergence year 2019).}
  \label{fig:gru-minority}
\end{figure}

At $k{=}4$ the additional cluster holds $802$ technologies ($0.40\%$). Its members share a consistent shape (Figure~\ref{fig:gru-minority}): a median of $23$ reuses at year 20 with an interquartile range of only $22$--$28$, reaching half of the final total by year $2$ and $90\%$ by year $3$ on average. But $91.9\%$ of them emerged in 2013 or later (median year 2019), so the cluster cannot be distinguished from an artifact of the 2022 data horizon: it subdivides the already censoring-dominated S1 rather than revealing a new long-run reuse behavior. Retaining $k{=}3$ keeps the horizon-shaped technologies in a single cluster while preserving the two growth patterns that are unambiguous under fuller observation.

\paragraph{The sequence-based labels are predictable from emergence-time features.} The sequence-based labels are derived from the trajectories alone, so predicting them from the seven emergence-time features is a prediction task with a genuinely external target, analogous to the trajectory task of Section~\ref{sec:results-transfer}. GBDT on all seven features reaches a one-vs-rest macro ROC-AUC of $0.914$ ($86.8\%$ accuracy, macro-$F_1$ $0.78$). Macro-$F_1$ falls below accuracy because it weights the small S1 class equally with the other two, and S1 is the class the argmax rule handles least well: its precision is $0.85$ but its recall only $0.56$, for an $F_1$ of $0.67$, against $0.76$ for S0 and $0.92$ for S2. This is a property of the decision threshold rather than a limit on how well S1 can be identified, since S1 is in fact the best-ranked of the three classes (one-vs-rest ROC-AUC $0.948$, and average precision $0.70$ against a $2.9\%$ base rate).

The figure is not an artifact of a single run or a single model. It is stable across three train/test split seeds with the labeling held fixed ($0.913$--$0.914$), ExtraTrees reaches $0.898$, and the FAE $K{=}3$ subset reaches $0.809$, reproducing the all-seven-versus-selected gap of the trajectory task.

At the matched cluster count $k{=}4$, the sequence-based labels are more predictable than the trajectory-shape labels of Section~\ref{sec:results-transfer} ($0.912$ versus $0.831$). Both figures are single fixed-seed runs and the training protocols differ slightly, since the trajectory-task classifier reserves an early-stopping split that the tree ensembles here do not need (Section~\ref{sec:setup}), so we read the direction rather than the exact margin: clustering in the learned representation, which compresses trajectory variation, yields a partition that emergence-time features can anticipate more accurately.

Set beside the agreement figures, these results make the underlying pattern explicit. The two labelings built from trajectories agree substantially with each other, at an ARI of $0.40$ between the sequence-based labeling and trajectory-shape $k$-means at three clusters and $0.44$ against four. The emergence-profile labeling agrees with neither, at $0.034$ and $0.043$. The dividing line is therefore not the clustering algorithm or the cluster count but the space being clustered: labelings derived from how reuse unfolded converge on one another, and a labeling derived from emergence-time covariates does not join them, so each route has to be validated on its own terms.

\paragraph{The contribution of emergence cohort.} Later cohorts are observed over a shorter part of the twenty-year window, so a technology's trajectory shape is partly determined by when it emerged, and $79.3\%$ of S1 in particular emerged in 2013 or later. The sequence labels may therefore be predictable from the features partly because the features carry cohort information: \texttt{ACCESS\_SIZE} has a Spearman rank correlation of $0.87$ with emergence year (Section~\ref{sec:limitations}). We test this directly by changing the predictors and leaving everything else fixed. Emergence year on its own, as a single predictor, reaches $86.9\%$ accuracy and a one-vs-rest macro ROC-AUC of $0.874$, against $86.8\%$ and $0.914$ for the seven features. Dropping the two truncated features \texttt{ACCESS\_SIZE} and \texttt{ACCESS\_TREND} lowers the seven-feature result to $0.850$, and adding emergence year to the seven raises it to $0.930$. Emergence year alone therefore matches the seven features on accuracy, in fact marginally exceeding them ($86.9\%$ against $86.8\%$), and falls four points short on ROC-AUC and seven on macro-$F_1$. What the seven features add over knowing nothing but the year is better ranking and better handling of the minority class, not additional accuracy.

We draw two conclusions, and they point in different directions. The forecasting task is not compromised: emergence year is observable at emergence exactly as the seven features are, so a practitioner forecasting the reuse class of a new technology may use it, and the $0.914$ remains an honest out-of-sample figure for that task. The interpretation is qualified: the $0.914$ is not by itself evidence that these emergence-time features capture technological substance, because a single calendar variable recovers most of it.

\paragraph{Separating right-censoring from cohort composition.} The first candidate explanation is the observation window: later cohorts are seen over less of the twenty-year span, so their trajectory shape is partly fixed before any technology-specific fact enters. We can test that explanation directly, because truncating every trajectory to a common horizon costs no new data. Technologies emerging in 2012 or earlier, $179{,}459$ of the corpus, are all observed for ten full years, so repeating the whole pipeline on a ten-year horizon over that subcohort removes right-censoring by construction. Holding the horizon at ten years and varying only whether the censored post-2012 cohorts are included, one-vs-rest macro ROC-AUC moves from $0.690$ with them to $0.661$ without, and the year-only figure from $0.672$ to $0.621$. Right-censoring is therefore worth about three points of ROC-AUC here, not the bulk of the cohort effect; emergence year still recovers the labels at $0.621$ when no technology in the sample is censored at all, which points to genuine change in the composition of patenting over the period rather than to the horizon. The larger movement is the horizon itself: at ten years the latent clusters are markedly less separated than at twenty (sampled silhouette $0.22$ against $0.47$), and less separated clusters are less predictable, which accounts for most of the difference between $0.914$ and these figures. We report the comparison at a fixed horizon for that reason, since the twenty-year and ten-year runs cluster different objects and their absolute scores are not interchangeable.

What this does not settle is whether the same three-point margin would hold at the twenty-year horizon used elsewhere in this study, which cannot be tested on this corpus: only the 2002 and 2003 cohorts are observed for twenty full years, so emergence year takes just two values there, and \texttt{ACCESS\_SIZE} is zero throughout the first of them and undercounted in the second (Section~\ref{sec:limitations}). Right-censoring and feature truncation are entangled in a way these data cannot separate, and a corpus extending before 2002 and beyond 2022 would be needed to do so.

\section{Discussion}
\label{sec:discussion}

\subsection{Interpreting the Results}

\paragraph{Why FAE selects \texttt{SIM\_TECH}.} The \texttt{SIM\_TECH} feature takes four distinct values, which follows directly from its definition as a hierarchical IPC distance \citep{chen2025predicting}. Because it is so low in cardinality, it is highly compressible, and this is likely why FAE keeps it: FAE's objective rewards reconstructing the full feature set, and a feature that compresses well is cheap to reconstruct. The same property means it contributes little geometric variance, which is why silhouette, a distance-based criterion, benefits less from it than from a continuous feature. This tension explains the pattern in Table~\ref{tab:cluster-metrics}: the FAE $K{=}3$ subset ranks only eleventh under silhouette, yet first and second under Calinski--Harabasz and Davies--Bouldin, which reward variance ratio and compactness rather than pairwise distance.

\paragraph{On the near-saturated classification accuracy.} The classifiers in Table~\ref{tab:classification} are almost perfect because the cluster labels are produced by $k$-means on the same features the classifiers receive: the labeling and learning stages are not separated, so an expressive model can approximate the labeling function very closely and the accuracy saturates. This is a property of evaluating against cluster assignments rather than an external ground truth, not a defect of any one model, and it is exactly why we do not stop at Table~\ref{tab:classification}. The external validation in Section~\ref{sec:results-transfer} shows that an externally constructed trajectory-shape labeling agrees with our emergence-profile labeling only marginally above chance (ARI $0.03$--$0.04$ at matched and unmatched cluster counts), so the two evaluations measure different things, and the claims we rest on the emergence-profile route go no further than what a consistency check supports.

\paragraph{Feature selection is task-specific.} The external evaluation clarifies what FAE does and does not provide. FAE selects the subset that best reconstructs the seven emergence-time features, and that subset produces well-separated, interpretable clusters. But the selection is not made for trajectory prediction, and it retains only the technology-component features, omitting all three early-invention features. The controlled ablation (Table~\ref{tab:ablation}) shows what this costs on the trajectory task and how: the seven-feature advantage over the FAE subset is distributed and redundant, with \texttt{INVENT\_APPL} the largest single contributor but no feature individually necessary. Reconstruction and prediction are different objectives, and we treat them as such, without claiming that one reinforces the other. Even so, the reconstruction-optimal subset reaches $0.740$ on the prediction task, in the same range as the $0.728$ reported by \citet{chen2025predicting} on a different corpus and labeling, which is more than its objective promises.

\paragraph{The two labelings are both predictable, and they describe different objects.} The two labeling routes partition the corpus differently (ARI $\approx 0.04$ between them), yet each partition is recoverable by supervised learning: the emergence-profile labels near-perfectly, as expected given the shared features, and the sequence-based labels at a ROC-AUC of $0.914$ from features the labeling never saw. At matched cluster count, the GRU-latent labels are more predictable from emergence-time features than the trajectory-shape labels ($0.912$ versus $0.831$ at $k{=}4$; single-run figures, with the protocol caveats of Section~\ref{sec:results-gru}). This suggests that the learned representation discards trajectory noise that emergence-time features cannot anticipate and retains the shape structure that those features do predict. The two routes also agree on a qualitative point: each isolates a small cluster dominated by right-censored recent technologies (the early-plateau cluster of the emergence-profile route and S1 of the sequence route), which is best read as a property of the observation window rather than of technology behavior. We therefore treat the two routes as answering different questions rather than as competing answers to one. The emergence-profile labels describe what a technology looked like when it appeared; the sequence-based labels describe how its reuse then unfolded. Both are legitimate taxonomies of the same corpus, and the near-zero agreement between them is the evidence that they are not substitutes. We draw the methodological consequence explicitly: a study that clusters emergence-time covariates and calls the result a reuse-pattern taxonomy is naming something it has not measured.

\paragraph{Practical implications.} The classifiers studied here forecast a technology's reuse class from information available in its emergence year, before the multi-year reuse trajectory can be observed. Decisions about which radical innovation projects to pursue are made under the same constraint, and how managers make them varies with their position in the firm \citep{wilden2023selecting}. The trajectory-task results indicate that the emergence-time features carry information about the eventual reuse pattern, under the labeling and corpus used here, though on both external targets a large part of the comparable figure is attributable to emergence cohort rather than to the features themselves (Sections~\ref{sec:results-transfer} and~\ref{sec:results-gru}), so the practical value of these features over simply knowing when a technology appeared is narrower than the headline scores suggest.

\paragraph{Sample efficiency is really about GBDT.} The sample-efficiency picture is not a deep-versus-tree story. At $1\%$ training data, TabM reaches $99.1\%$ macro-$F_1$ against GBDT's $98.0\%$, roughly a $55\%$ reduction in error rate, but the tree-based ExtraTrees ($98.6\%$) also sits well above GBDT. What the sweep shows is that GBDT degrades faster than ExtraTrees, TabM, and FT-Transformer at very low data, and that the effect is confined to the smallest fractions. TabNet is a separate outlier: its accuracy falls off fastest of all at small training fractions. One possible explanation, which we did not test, is that its sparse attention mechanism is sensitive to the very small absolute number of minority-class examples at the smallest fractions: at $1\%$ of the training set, the early-plateau cluster contributes roughly $37$ of $1{,}089$ examples. One caveat applies to the whole sweep: it is run on the emergence-profile labels, where every model sits above $97\%$ macro-$F_1$ even at $1\%$ of the training data, so it compares models that all already solve the task, and the whole spread it resolves is under two points. Whether the same ordering holds on the harder sequence-based target, where macro-$F_1$ is $0.78$ at full data, we have not tested.

\paragraph{Interpretability differs across the model families.} The seven classifiers are not equally transparent, which matters for how a practitioner might use them. The tree ensembles (GBDT, ExtraTrees) and TabKAN are relatively interpretable: tree models expose their splits and feature importances, and TabKAN's Kolmogorov--Arnold layers learn explicit univariate functions, parameterized by Chebyshev coefficients, that can be written out per feature. The remaining deep models (TabNet, FT-Transformer, TabM, TabMixer) are effectively black boxes, producing predictions without a directly readable account of how they were reached. Our application does not demand interpretability, since the goal is an accurate early forecast rather than a decision that must be justified to a regulator. In domains such as medicine or law, where a prediction has to be explained, the interpretable members of this benchmark would be preferable even at some cost in accuracy.

\subsection{Limitations}
\label{sec:limitations}

This study has six limitations.

First, the evaluation is single-domain: it uses USPTO patents from 2002--2022, and we have not tested whether the findings transfer to other patent offices such as the EPO or JPO.

Second, the cluster labels are unsupervised proxies rather than external ground truth; the independently constructed trajectory-shape labeling in Section~\ref{sec:results-transfer} reduces this concern but does not remove it.

Third, FAE's selection is not stable under data resampling. Refitting FAE on five bootstrap subsamples at each of three subsample fractions ($50\%$, $75\%$, $90\%$), the mean pairwise Jaccard similarity of the selected $K{=}3$ subsets ranges from $0.28$ to $0.43$, below the conventional $0.5$ threshold, and no single feature is selected in more than $80\%$ of trials; at $K{=}5$ the mean Jaccard rises to $0.52$. This is a stricter test than the smoothness-across-$K$ criterion of the original FAE evaluation \citep{wu2021fractal}, which our $K$-sweep does satisfy, but the instability should be kept in mind wherever selection consistency itself is the quantity of interest.

Fourth, the corpus is right-censored and success-conditioned. Only $34.2\%$ of technologies are observed for the full 20-year window, and trajectories of later cohorts are flat by construction beyond the 2022 data horizon; this particularly affects the small early-plateau cluster, whose members are predominantly recent (Section~\ref{sec:results-clustering}). Any labels built from right-censored trajectories, including the trajectory-shape labels, inherit the observation window of the data they are built on. Within the window, a delayed take-off of the kind documented for sleeping-beauty patents \citep{hou2019patent} cannot be distinguished from censoring. The 20-reuse filter additionally conditions the corpus on substantial reuse, so our labels describe the pattern of reuse among technologies that reached that threshold and say nothing about whether a technology will reach it.

Fifth, two of the seven features are left-truncated for the earliest cohorts. \texttt{ACCESS\_SIZE} and \texttt{ACCESS\_TREND} are computed over the five years before a technology emerges, but the patent record we assembled begins in 2002, so for technologies emerging in 2002 that window lies entirely outside the data and \texttt{ACCESS\_SIZE} is zero for all $26{,}471$ of them; it remains undercounted until the first fully covered cohort in 2007. Affected cohorts, 2002 to 2006, are $137{,}118$ of $201{,}710$ technologies, or $68\%$ of the corpus, and \texttt{ACCESS\_SIZE} consequently rises steeply with emergence year across that range (Spearman $\rho = 0.87$ over the whole corpus).

Because \texttt{ACCESS\_SIZE} is one of the three features FAE selects, part of what the emergence-profile partition separates is emergence cohort rather than technological position, and the cluster contrasts of Section~\ref{sec:results-clustering} are correspondingly overstated: restricted to the fully covered 2007 and later cohorts, the early-plateau cluster's component-access advantage over the other two falls from a factor of fourteen to twenty-nine down to six to ten. The ordering of the three clusters on this feature is unchanged under that restriction, but its magnitude is not, and comparisons involving \texttt{ACCESS\_SIZE} or \texttt{ACCESS\_TREND} across cohorts of different ages should be read as upper estimates.

The truncation does not touch the construction of the sequence-based or trajectory-shape labelings, which are computed from the trajectories alone, but the classifiers that predict them take the truncated features as inputs, and Section~\ref{sec:results-gru} reports what that is worth: emergence year by itself recovers the sequence labels at ROC-AUC $0.874$ against $0.914$ for the seven features, and removing the two truncated features drops the seven-feature figure to $0.850$. Right-censoring and feature truncation therefore compound, the first shaping the labels and the second supplying a predictor correlated with cohort.

Section~\ref{sec:results-gru} separates the two as far as these data allow, by repeating the pipeline at a ten-year horizon on the $179{,}459$ technologies that are all observed for that full span: right-censoring is worth about three points of ROC-AUC there. The same test cannot be run at the twenty-year horizon, because only the 2002 and 2003 cohorts are fully observed, and \texttt{ACCESS\_SIZE} is zero or undercounted throughout them. Extending the patent record back to the mid-1990s would remove the truncation, and would also let the first-ever co-occurrence that defines a novel technology be verified rather than assumed for the 2002 cohort; that history was not available to us, and without it the two effects cannot be fully disentangled.

Sixth, the trajectory-task scores (Table~\ref{tab:ablation}, Figures~\ref{fig:fae-vs-all7} and~\ref{fig:feature-budget}) are single runs at a fixed seed, as is Table~\ref{tab:classification}; only the sample-efficiency sweep averages over multiple seeds. The margins we interpret are large ($0.740$ versus $0.831$), but small differences between models in these tables should not be over-read.

\section{Conclusion}
We studied whether the long-term reuse of a newly emerged patent technology can be predicted from information available at its emergence. From a large corpus of USPTO patents we built a set of $201{,}710$ novel technologies. Because no reuse-pattern labels exist in advance, we constructed them, and the construction turned out to matter as much as the prediction.

Our primary labeling clusters the reuse trajectories in the latent space of a GRU autoencoder trained on those trajectories alone; emergence-time features recover it at a one-vs-rest macro ROC-AUC of $0.914$, and at a comparable level under the alternative cluster count. That figure carries one qualification: the calendar year of emergence recovers the same labels on its own at $0.874$. A replication on technologies observed for ten full years, none of them right-censored, indicates that this mainly reflects change over time in what was patented rather than the shorter observation of recent cohorts. The forecast stands, since emergence year is known when a technology appears, but the margin attributable to the seven features is roughly four points of ROC-AUC rather than the whole of it.

A second route, Fractal Autoencoder feature selection followed by $k$-means on the selected emergence-time features, yields a partition that classifiers recover near-perfectly but that agrees with the trajectory-based partition barely above chance. We read that gap as a substantive result in its own right: clustering emergence-time features produces a taxonomy of technologies as they begin, which is a different object from a taxonomy of how they go on to be reused, and the two are not interchangeable even though they are often treated as one.

Retrained on a trajectory-shape task built to follow the published construction, the models reach $0.821$ to $0.831$ once all seven features are used (GBDT $0.831$); the $0.728$ reported by \citet{chen2025predicting} on a different corpus and labeling serves as a reference point rather than a benchmark, and a controlled ablation shows the seven-feature advantage is distributed across the features rather than owed to any single one. Both routes independently isolate a small cluster of right-censored recent technologies, which underlines that any trajectory labeling inherits the observation window that produced it.

Several extensions follow naturally. Censoring-aware training, for example masking the autoencoder's reconstruction loss beyond each technology's observed horizon, would separate genuine saturation from the data horizon. Evaluations on other patent offices would test whether the reuse patterns and their predictability transfer beyond the USPTO. More broadly, the combination of unsupervised label construction, supervised validation, and learned sequence representations is not specific to patents and could be applied wherever labels must be constructed before prediction is possible.


\section*{Declarations}

\noindent\textbf{Funding.} This research is supported in part by the NSF under grants IIS 2327113 (with a supplement from the CISE Research Experience for Undergraduates Student Funding Program, administered by the Computing Research Association), DMS-2208314, and ITE 2433190, as well as by the NIH under grants R21AG070909 and P30AG072946.

\medskip

\noindent\textbf{Competing interests.} The authors declare that they have no competing interests.

\medskip

\noindent\textbf{Data availability and code availability.} The patent records analyzed in this study are publicly available from the United States Patent and Trademark Office (USPTO) Open Data Portal and were retrieved through its Open Data Portal API. The analysis code, together with the scripts and documented steps needed to rebuild the technology-level dataset from those records and to reproduce every reported result, is available at \url{https://github.com/ayhamyousef/innovation_prediction}.

\medskip

\noindent\textbf{Author contributions.} \textbf{Ayham Yousef:} Methodology, Formal analysis, Software, Investigation, Writing -- original draft, Writing -- review \& editing. \textbf{Qiang Ye:} Methodology, Investigation, Supervision, Project administration, Resources, Funding acquisition, Writing -- review \& editing. \textbf{Qiang Cheng:} Conceptualization, Methodology, Investigation, Project administration, Funding acquisition, Writing -- review \& editing.

\bibliographystyle{plainnat}
\bibliography{references}

\clearpage \appendix
\section*{Appendices}
\section{Classifier Hyperparameters}
\label{app:hyperparams}

Table~\ref{tab:hyperparams} lists the hyperparameters used for each classifier. All deep tabular models are trained with the AdamW optimizer, a batch size of $256$, a maximum of $200$ epochs, and early stopping with patience $20$ on the validation split; TabNet uses the optimizer of its reference implementation. The tree ensembles use the scikit-learn implementations with the settings shown. All models use a fixed random seed of $42$.

\begin{table}[htbp]
\centering
\caption{Hyperparameters by classifier.}
\label{tab:hyperparams}
\begin{tabular}{@{}ll@{}}
\toprule
Model & Settings \\
\midrule
GBDT           & $200$ estimators, max depth $5$, learning rate $0.1$ \\
ExtraTrees     & $100$ estimators \\
TabNet         & $n_d{=}n_a{=}16$, $5$ steps, $\gamma{=}1.5$, lr $0.02$ \\
FT-Transformer & token dim $64$, $3$ blocks, $4$ heads, attn.\ dropout $0.2$, lr $10^{-4}$ \\
TabM           & $8$-member ensemble, hidden $[128, 64]$, dropout $0.1$, lr $10^{-3}$ \\
TabKAN         & Chebyshev-KAN mixer, $4$ layers, token/channel dim $64$/$128$, degree $3$, lr $10^{-3}$ \\
TabMixer       & feature dim $64$, feed-forward dim $256$, lr $10^{-3}$ \\
\bottomrule
\end{tabular}
\end{table}

For the deep models, the weight decay is $10^{-5}$. Features are standardized using training-set statistics before being passed to any classifier.

\section{Misclassified Test Technologies}
\label{app:errors}

\begin{figure}[htbp]
  \centering
  \includegraphics[width=\linewidth]{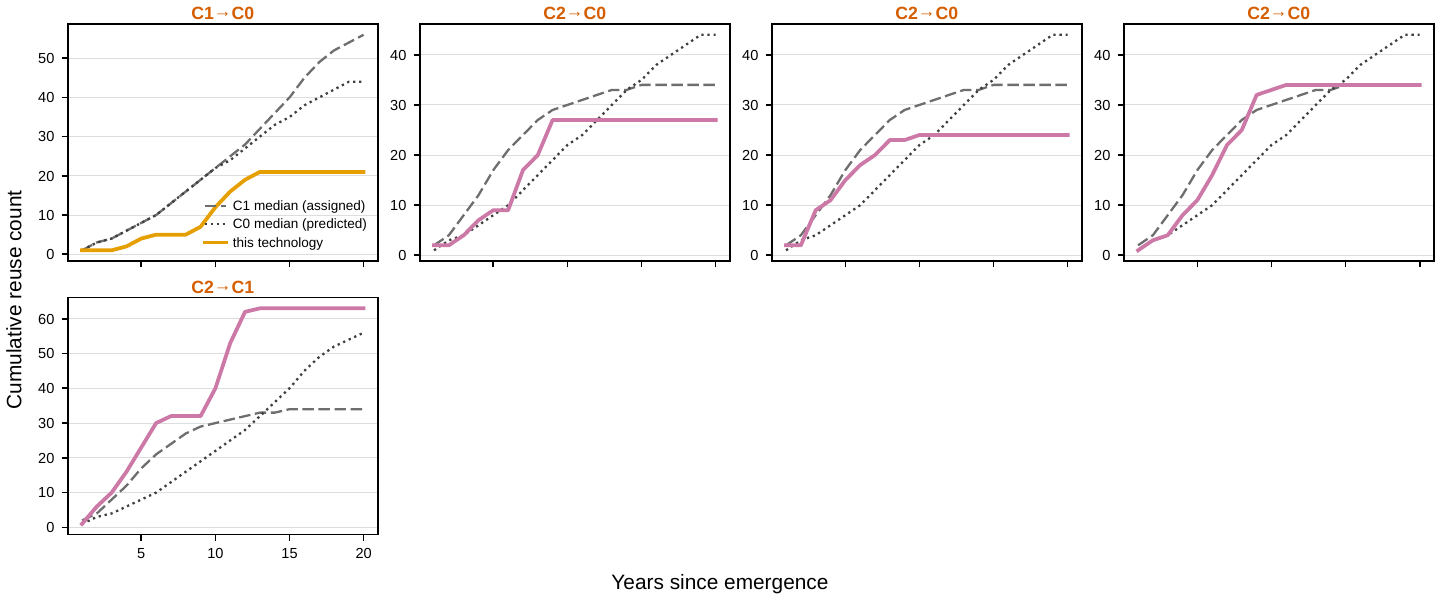}
\caption{All misclassified test technologies ($5$ of $80{,}684$). Each is a boundary case whose trajectory is atypical for its assigned cluster. The solid line is the technology, the dashed and dotted lines are the assigned- and predicted-cluster medians. Color denotes the assigned (clustering) label.}
  \label{fig:casestudy-err}
\end{figure}

Figure~\ref{fig:casestudy-err} shows all five technologies (of $80{,}684$ in the test set) misclassified by TabMixer, discussed in Section~\ref{sec:results-casestudy}. Each is a boundary case whose trajectory is atypical for its assigned cluster.

\section{Sample-Efficiency Values}
\label{app:fraction}

Table~\ref{tab:fraction} gives the exact macro-$F_1$ values plotted in Figure~\ref{fig:fraction}, for readers who want the underlying numbers.

\begin{table}[htbp]
\centering
\caption{Test macro-$F_1$ (\%) by training-set fraction on FAE $K{=}3$ features (mean over three seeds). $n$ is the number of training examples.}
\label{tab:fraction}
\setlength{\tabcolsep}{4pt}
\begin{tabular}{@{}lrrrrr@{}}
\toprule
Fraction ($n$) & GBDT & TabM & FT-Tr. & ExtraTr. & TabNet \\
\midrule
0.01 ($1{,}089$)    & 98.0 & 99.1 & 99.0 & 98.6 & 97.2 \\
0.05 ($5{,}445$)    & 99.1 & 99.4 & 99.3 & 99.5 & 98.4 \\
0.10 ($10{,}891$)   & 99.6 & 99.7 & 99.7 & 99.6 & 99.0 \\
0.25 ($27{,}230$)   & 99.6 & 99.9 & 99.8 & 99.7 & 99.1 \\
0.50 ($54{,}461$)   & 99.7 & 99.9 & 99.9 & 99.8 & 99.3 \\
1.00 ($108{,}923$)  & 99.8 & 99.9 & 99.9 & 99.9 & 99.4 \\
\bottomrule
\end{tabular}
\end{table}

\end{document}